\documentclass[preprint,3p]{elsarticle}

\journal{}
\usepackage{graphicx}
\usepackage{dcolumn}
\usepackage{bm}
\usepackage{bigints}
\usepackage[mathlines]{lineno}
\usepackage{comment}
\linenumbers\relax 

\usepackage[utf8]{inputenc}
\usepackage[T1]{fontenc}
\usepackage{amsmath,amssymb}
\usepackage{etoolbox}
\usepackage{bbold}
\usepackage{color}
\usepackage{pgfplots}
\pgfplotsset{compat=newest}
\usepackage{tikz}
\usetikzlibrary{patterns}

\makeatletter
\def\thickhline{%
  \noalign{\ifnum0=`}\fi\hrule \@height \thickarrayrulewidth \futurelet
   \reserved@a\@xthickhline}
\def\@xthickhline{\ifx\reserved@a\thickhline
               \vskip\doublerulesep
               \vskip-\thickarrayrulewidth
             \fi
      \ifnum0=`{\fi}}
\makeatother

\newlength{\thickarrayrulewidth}
\journal{Journal of Sound and Vibration}

\begin{document}
\nolinenumbers

\begin{frontmatter}

\title{New approach to approximate analytical solutions for a harmonic oscillator subject to combined linear and nonlinear damping}
\author[1]{Karlo Lelas\corref{cor1}}
\ead{klelas@ttf.unizg.hr}

\author[2]{Robert Pezer}

\affiliation[1]{
organization={Faculty of Textile Technology, University of Zagreb, Prilaz baruna Filipovića 28a, 10000 Zagreb},
country={Croatia}
}
\affiliation[2]{
organization={Department of Physical Metallurgy, Faculty of Metallurgy, University of Zagreb, Aleja narodnih heroja 3, 44000 Sisak},
country={Croatia}
}
\cortext[cor1]{Corresponding author}


\begin{abstract}
We introduce a new approach to deriving approximate analytical solutions of a harmonic oscillator damped by purely nonlinear, or combinations of linear and nonlinear damping forces. Our approach is based on choosing a suitable trial solution, i.e. an ansatz, which is the product of the time-dependent amplitude and the oscillatory (trigonometric) function with a reduced frequency (due solely to linear damping) and an initial phase shift (due to both linear and nonlinear damping) compared to the undamped case. We derive the amplitude decay equation using the connection of the energy dissipation rate with the power of the total damping force. By matching our ansatz to the initial conditions, we obtain equations for its initial amplitude and initial phase. We demonstrate the validity of our approach for quadratic damping, a combination of linear and quadratic damping, and finally a combination of linear, quadratic, and Coulomb damping. For pure quadratic damping and for a combination of linear and quadratic damping, we compare our approximate analytical solutions with the corresponding numerical solutions and find that they describe the oscillator dynamics well in the regime of weak to moderately strong damping. For the combination of all three types of damping, our solutions agree well with the numerical solutions until the last half-period or the last few half-periods of the motion (primarily depending on the strength of the Coulomb damping). In addition, we show how to use our solutions to determine damping constants from given displacement vs. time data, demonstrating that they can be applied to the analysis of experiments.
\end{abstract}


\end{frontmatter}

\section{Introduction}

In physics and engineering, the decrease in the amplitude of free harmonic oscillations is usually attributed to one of three types of damping: damping linear in velocity, damping quadratic in velocity, and sliding friction. In the case of a harmonic oscillator damped by a force linear in velocity (viscous damping), the corresponding equation of motion can be solved exactly, and the solutions can be easily analyzed \cite{Waves, Cutnell8, Resnick10, Young2020university}. In case of a harmonic oscillator damped with sliding friction (Coulomb damping), the corresponding equation of motion is nonlinear because the friction force reverses direction with velocity, which is modeled via the signum function, but it can still be solved exactly by splitting the motion into left and right-moving segments \cite{Lapidus, AviAJP, Grk2, Kamela}, i.e. the motion needs to be analyzed over half-periods and the solution thus obtained is piecewise defined. For a harmonic oscillator with damping quadratic in velocity, the corresponding equation of motion is a nonlinear differential equation that cannot be solved analytically, so approximate or numerical methods must be used \cite{Smith, Mungan, Wang, Grk2}. Therefore, even for a harmonic oscillator with only one type of damping, the dynamics can be challenging to analyze if the damping is nonlinear, i.e. Coulomb damping or damping quadratic in velocity.

Although the damped harmonic oscillator model, with a single type of damping, can successfully describe some simple real-world systems, such as physical pendulums \cite{Wang}, block-spring systems \cite{Kamela}, RLC circuits \cite{Lelas2023}, etc., for a more complete and precise description, it is necessary to take into account that oscillations are damped simultaneously by two or all three types of damping. For example, in \cite{AJPpendulum} a simple physical pendulum was studied, both experimentally and theoretically, and it was shown that the contribution of all three types of damping must be taken into account for an adequate general description of pendulum damping, while in some specific cases two types of damping were sufficient. Many authors have investigated damping of harmonic oscillations by the combined action of two types of damping. For free vibrations with combined Coulomb and linear damping, the amplitude decay was first derived under no-stop conditions in \cite{Markho}, while the theory of amplitude decay presented in \cite{AJPRicch} accounts for the stopping of dynamics with an offset from the equilibrium position. The exact theory of discontinuities in the acceleration of such systems is given and experimentally verified in \cite{AJPHinrich}. Although this specific combination of damping allows exact analytical analysis, the solutions are piecewise defined, and simpler closed form approximate analytical solutions remain a subject of interest \cite{JSVCoulombViscous}. For a combination of a damping quadratic in velocity and any other type of damping, an exact approach to solving the corresponding equation of motion is not possible. For this case, the approximate analytical expression for the amplitude decay was derived using the work energy theorem already in \cite{AJPNelson}, while numerical methods are routinely used; e.g., see \cite{Bacon}.  

In a recent paper \cite{LelasPezer2}, a harmonic oscillator damped simultaneously by all three types of damping was studied, and approximate analytical solutions were found that agree excellently with numerical solutions, but only in the weak damping regime. The approach used in \cite{LelasPezer2} is based on ansatz of the form $\tilde{x}(t)=A_0\tilde{f}(t)\cos(\omega_0t+\varphi_0)$, where $\omega_0$, $A_0$, and $\varphi_0$ are the angular frequency, amplitude, and initial phase of the corresponding undamped solution, while $\tilde{f}(t)$ is the unknown function that describes the decay of the amplitude over time. The connection of the energy dissipation rate and the power of the total damping force was used to determine the unknown function $\tilde{f}(t)$ \cite{LelasPezer2}. More precisely, in the case of weak damping, one can take that $\tilde{f}(t)$ remains approximately constant over half cycles (half periods) and average the oscillatory parts of the energy dissipation rate to obtain the first-order differential equation for $\tilde{f}(t)$. The same approach was previously used to determine amplitude decay for each of the three damping types individually \cite{Wang}, for the combination of linear and quadratic damping \cite{AJPNelson}, and for the combination of Coulomb and linear damping \cite{Vitorino}. 
This approach is mathematically simple, but as damping increases, the angular frequency, initial phase, and initial amplitude can change significantly compared to the undamped case. 

The analysis of the influence of the damping strength on the initial phase and frequency, carried out in \cite{LelasPezer} for linear damping and in \cite{LelasPezer2} for quadratic damping, clearly demonstrates that increasing the strength of linear and quadratic damping leads to significant changes in the initial phase and initial amplitude compared to the undamped case. Unlike linear damping, which noticeably reduces the oscillation frequency even in the weak regime \cite{LelasPezer}, quadratic damping has been shown to have a negligible effect on the frequency of harmonic oscillations, at least in the weak to moderate damping regime \cite{LelasPezer2}. Therefore, here we use an ansatz of the form $x(t)=Af(t)\cos(\omega t+\varphi)$ as the basis of our approach, with $\omega=\omega_0$ in the case of pure nonlinear damping (e.g., for a harmonic oscillator with quadratic damping only), but for combinations of linear and nonlinear damping forces, we take $\omega<\omega_0$, where $\omega$ is given by the exact analytical formula for harmonic oscillations with linear damping. Thus, in addition to changing the initial amplitude and phase, we also use the fact that linear damping reduces the frequency. In this approach, besides the unknown function $f(t)$, we have to determine the constants $A$ and $\varphi$, i.e. we have a somewhat more complex mathematical problem compared to the approach used in \cite{LelasPezer2}, but, as we will show, we obtain approximate analytical solutions that describe the dynamics significantly better and are valid for a wider range of values of the damping constants than the solutions obtained in \cite{LelasPezer2}. 

Regarding the effect of Coulomb damping on a harmonic oscillator, it leaves the time intervals between successive turning points the same as those of an undamped oscillator and, in that sense, does not alter its frequency, but it introduces increasing time shifts of the zero crossings relative to those of the undamped case; furthermore, it ultimately leads to the cessation of motion at some finite displacement (i.e. out of equilibrium) \cite{LelasPezer2}. We note here that a common feature of the new approximate analytical solutions and those obtained in \cite{LelasPezer2} is that, due to their very form, they cannot model with high precision the effects arising from Coulomb damping at the very end of the dynamics, a matter that we discuss in detail later in the paper.

The rest of the paper is organized into six sections. In Section \ref{Basic}, we consider the case of harmonic oscillator damped simultaneously by three types of damping, introduce the theory and approximations we use, derive the first-order differential equation for the amplitude decay, the equation for the initial amplitude, and the equation for the initial phase. In Section \ref{quadratic}, we obtain new approximate analytical solutions for a harmonic oscillator with quadratic damping. 
In Section \ref{LinearQuadratic}, we obtain new approximate analytical solutions for combined linear and quadratic damping. In Section \ref{AllThree} we obtain new approximate analytical solutions for the combination of all three types of damping. In Sections \ref{quadratic}, \ref{LinearQuadratic}, and \ref{AllThree}, we compare the new solutions with the corresponding solutions obtained by the approach used in \cite{LelasPezer2} and with numerical solutions. We note here for clarity that throughout the paper new solutions are designated with $x(t)$, corresponding energies with $E(t)$, and both are shown as solid red curves in all figures. The solutions obtained by the approach used in \cite{LelasPezer2} are designated with $\tilde{x}(t)$, corresponding energies with $\tilde{E}(t)$, and both are shown as solid blue curves in all figures, while numerical solutions and numerically obtained energies are shown as dotted black curves in all figures.
We compare $x(t)$ and $\tilde{x}(t)$ throughout this work to highlight the importance of accounting for changes in the initial amplitude, initial phase, and frequency (due solely to linear damping), relative to the undamped case, in order to achieve good agreement with the numerical results. At the end of each of Sections \ref{quadratic}, \ref{LinearQuadratic}, and \ref{AllThree} we comment on the results for the type of damping considered. In Section \ref{application} we show how to use our approximate analytical solutions to determine the damping constants from given displacement vs. time data. Finally, in Section \ref{Conclusion}, we summarize the key findings of the paper and provide an outlook.

\section{New equations for the amplitude decay, initial amplitude, and initial phase}
\label{Basic}

\begin{figure}[h!t!]
\begin{center}
\includegraphics[width=0.6\textwidth]{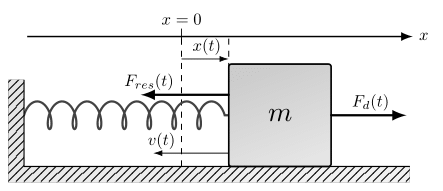}
\end{center}
\caption{Schematic representation of a block-spring system with a restoring force $F_{res}(t)$ and a (total) damping force $F_d(t)$. Here we show the time instant at which $x(t)>0$ and $v(t)<0$, i.e. at which $F_{res}(t)<0$ and $F_d(t)>0$. See text for details.} 
\label{shema}
\end{figure}

We consider the paradigmatic example of a damped harmonic oscillator shown in Fig.\,\ref{shema}, i.e. a block of mass $m$ that oscillates back and forth along a horizontal surface under the influence of the restoring force of an ideal spring $F_{res}(t)=-kx(t)$, where $k$ is the stiffness of the spring and $x(t)$ is the displacement of the block from the equilibrium position (set to $x=0$), and under the influence of the damping force
\begin{equation}
F_d(t)=-\text{sgn}[v(t)]\mu mg-bv(t)-Dv(t)|v(t)|\,,
\label{Fd}
\end{equation}
where $\mu>0$, $b>0$ and $D>0$ are the corresponding damping constants, $v(t)=dx(t)/dt$ is the velocity, and  
\begin{equation} \label{signum}
    \text{sgn}\left[v(t)\right] = \begin{cases}
\begin{tabular}{@{}cl@{}}
   $1$\, & if\, $v(t)$ $ > $ $0$ \\
    $0$\, & if\, $v(t)$ $=$ $0$ \\
    $-1$\, & if\, $v(t)$ $<$ $0$\,
\end{tabular}
    \end{cases}
    \end{equation}
is the sign function. The corresponding equation of motion is
\begin{equation}
ma(t)=-\text{sgn}\left[v(t)\right]\mu mg-bv(t)-Dv(t)|v(t)|-kx(t)\,,
\label{HOeq}
\end{equation}
where $a(t)=d^2x(t)/dt^2$ is the acceleration of the block. The first term on the right-hand side of equation \eqref{HOeq} models the force of sliding friction \cite{Lapidus, AviAJP, Grk2, AJPHinrich, JSVCoulombViscous}, while the second and third terms model, e.g., the influence of air resistance, since air resistance generally depends on terms proportional to velocity and square of velocity \cite{AJPpendulum, AJPNelson, Bacon, Wang}. For simplicity, here we assume that the static and dynamic coefficients of friction are the same, i.e. both equal $\mu$. Consequently, if the block comes to rest $(v = 0)$ at a position where $|kx|\leq\mu mg$, static friction balances the spring force and the block remains at rest; otherwise, it starts moving again in the direction of the spring force. The numerical solutions in this paper account for this sticking condition; i.e. the motion terminates at the first turning point where $|kx|\leq\mu mg$ holds. 
The energy (potential plus kinetic) of the block-spring system is given by  
\begin{equation}
E(t)=\frac{kx^2(t)}{2}+\frac{mv^2(t)}{2}\,.
\label{Energy}
\end{equation}
If we put $\mu=b=D=0$ in \eqref{HOeq}, i.e. for $F_d(t)=0$, we get the equation of the undamped harmonic oscillator \cite{Resnick10}, with a general solution 
\begin{equation}
x_{HO}(t)=A_0\cos(\omega_0t+\varphi_0)\,,
\label{xHO}
\end{equation}
where $\omega_0=\sqrt{k/m}$ is the angular frequency of the undamped system, while $A_0$ and $\varphi_0$ are constants (amplitude and initial phase) that are determined from initial conditions $\left(x_0,v_0\right)$, i.e. 
\begin{equation}
A_0=\sqrt{x_0^2+\left(\frac{v_0}{\omega_0}\right)^2}\,\,\,\text{and}\,\,\,\varphi_0=\arctan\left(-\frac{v_0}{\omega_0x_0}\right)\,.
\label{A0phi0}
\end{equation}
%
The undamped system oscillates with conserved energy \cite{Resnick10}. For damped systems, the energy is not conserved due to the power of the damping force $P_d(t)=F_d(t)v(t)$, and the energy dissipation rate is given by $dE(t)/dt=P_d(t)$, i.e. 
\begin{equation}
\frac{dE(t)}{dt}=-\mu mg|v(t)|-bv^2(t)-D|v(t)|^3\,,
\label{Power}
\end{equation}
where we used $\text{sgn}\left[v(t)\right]v(t)=|v(t)|$ and $v^2(t)|v(t)|=|v(t)|^3$. Thus, equation \eqref{Power} tells us that the energy of the damped system decreases monotonically with time, i.e. $\frac{dE(t)}{dt}\leq0$ for all $t\geq0$, and $\frac{dE(t)}{dt}=0$ holds at the turning points, i.e. at instants when $v(t)=0$. For later convenience, we note here that one can take a moving average of the relation \eqref{Power} and obtain
\begin{equation}
\frac{d\langle E(t)\rangle}{dt}=-\mu mg\langle|v(t)|\rangle-b\langle v^2(t)\rangle-D\langle|v(t)|^3\rangle\,,
\label{PowerAv}
\end{equation}
where we use the notation $\langle F(t)\rangle=(\Delta T)^{-1}\int_t^{t+\Delta T}F(t')dt'$. The relation \eqref{PowerAv} is exact, provided that $\Delta T$ is constant.

In a recent paper \cite{LelasPezer2}, approximate analytical solutions of the harmonic oscillator weakly damped by a force \eqref{Fd} were derived starting from the simple ansatz
\begin{equation}
\tilde{x}(t)=A_0\tilde{f}(t)\cos(\omega_0t+\varphi_0)\,,
\label{Xansatz0}
\end{equation}
where $A_0$, $\omega_0$ and $\varphi_0$ are the same as in the undamped case, i.e. as in \eqref{xHO}, while the unknown function $\tilde{f}(t)$ describes the slow decay of oscillations over time. By slow decay we mean that the rate of change of the function $\tilde{f}(t)$ over time is significantly slower than the rate of change of the function $\cos(\omega_0t+\varphi_0)$. The approach in \cite{LelasPezer2} is further based on taking
\begin{equation}
\tilde{v}(t)=-\omega_0A_0\tilde{f}(t)\sin(\omega_0t+\varphi_0)\,
\label{Vansatz0}
\end{equation}
for velocity, i.e. in neglecting the term with $d\tilde{f}(t)/dt$ in $d\tilde{x}(t)/dt$, which leads to energy of the form
\begin{equation}
\tilde{E}(t)=\frac{m\omega_0^2A_0^2\tilde{f}^2(t)}{2}\,.
\label{Eansatz0}
\end{equation}
These steps are justified in the case of weak damping, i.e. when the frequency and initial phase remain approximately the same as in the undamped case \cite{LelasPezer2}. 
A simple first order differential equation for the function $\tilde{f}(t)$ is obtained by inserting velocity \eqref{Vansatz0} and energy \eqref{Eansatz0} into the energy dissipation rate \eqref{Power} and taking the time average of trigonometric functions over half periods \cite{LelasPezer2}. It was shown (for a particular choice of $m$, $k$ and $A_0$) that the approximate analytical solutions obtained by this approach are in excellent agreement with the numerical solutions for a certain limited range of values of the damping constants ($\mu$, $b$ and $D$) and that this agreement also depends on the type of initial conditions considered (i.e. on $\varphi_0$) \cite{LelasPezer2}. In particular, the agreement of approximate analytical solutions of the form \eqref{Xansatz0} with numerical solutions is better in the case of initial conditions $(x_0=0, v_0\neq0)$ with a purely kinetic initial energy than in the case of initial conditions $(x_0\neq0,v_0=0)$ with a purely potential initial energy \cite{LelasPezer2}.   

Here we improve the approach used in \cite{LelasPezer2} by starting from the ansatz of the form
\begin{equation}
x(t)=Af(t)\cos(\omega t+\varphi)\,,\,\text{with}\,\,\omega=\sqrt{\omega_0^2-d_1^2}\,,
\label{Xansatz1}
\end{equation}
where $d_1=b/(2m)$, $A\neq A_0$ and $\varphi\neq\varphi_0$ are unknown constants, and $f(t)$ is the unknown function that describes the amplitude decay over time. The corresponding velocity is
\begin{equation}
v(t)=\frac{dx(t)}{dt}=-\omega Af(t)\sin(\omega t+\varphi)+A\frac{df(t)}{dt}\cos(\omega t+\varphi)\,.
\label{Vansatz1}
\end{equation}
Although we have not yet determined the function $f(t)$, without loss of generality, we can take $f(0)=1$, and using initial conditions $(x(0)=x_0, v(0)=v_0)$ we obtain the equations
\begin{equation}
x_0=A\cos(\varphi)\,
\label{pocuvjet1}
\end{equation}
and
\begin{equation}
v_0=-\omega A\sin(\varphi)+A\frac{df(t)}{dt}\bigg|_{t=0}\cos(\varphi)\,
\label{pocuvjet2}
\end{equation}
from which we will determine the constants $A$ and $\varphi$ once we determine $df(t)/dt$ at $t=0$. The energy corresponding to \eqref{Xansatz1} and \eqref{Vansatz1} is
\begin{equation}
\begin{split}
E(t) &=\frac{m\omega_0^2A^2}{2}f^2(t)\left[1-\left(\frac{d_1}{\omega_0}\right)^2\sin^2(\omega t+\varphi)-\right.\\
     &\quad\left.-\left(\frac{\omega}{\omega_0}\right)\left(\frac{1}{\omega_0 f(t)}\frac{df(t)}{dt}\right)\sin(2\omega t+2\varphi)+\left(\frac{1}{\omega_0 f(t)}\frac{df(t)}{dt}\right)^2\cos^2(\omega t+\varphi)\right]\,.
\end{split}
\label{energija}
\end{equation}
Written in this form, the energy shows that, relative to the leading term $m\omega_0^2A^2f^2(t)/2$, the remaining terms are of the order $(d_1/\omega_0)^2$, $\beta(t)$, and $\beta^2(t)$, where
\begin{equation}
\beta(t)=\frac{1}{\omega_0 f(t)}\left|\frac{df(t)}{dt}\right|
\label{beta}
\end{equation}
is the relative decay rate of the amplitude per radian of the oscillation. The quantity $(\omega_0/\omega)\beta(t)$ measures the ratio of the two terms in the velocity \eqref{Vansatz1}. We note that keeping the term with $df(t)/dt$ in the velocity \eqref{Vansatz1} is necessary to ensure that our modeled displacement \eqref{Xansatz1} and velocity \eqref{Vansatz1} can be matched to the initial conditions. Correspondingly, we must keep all terms in \eqref{energija}, regardless of the damping strength, to ensure that the modeled energy has the correct initial value. Thus, keeping all these terms ensures that the modeled displacement, velocity, and energy satisfy the initial conditions for any function $f(t)$ we use to model the amplitude decay, as long as $f(0)=1$ and $df(t)/dt$ has a finite value at $t=0$.

When modeling the function $f(t)$ itself, we can be more flexible with these terms, and use the fact that they are negligible in the weak damping limit, which we define by the condition
\begin{equation}
\beta(t)\ll 1\,.
\label{wdcond}
\end{equation}
We stress that the condition must be imposed on the relative decay rate $\beta(t)$, not on $\omega_0^{-1}|df(t)/dt|$ alone, because $f(t)$ itself decreases with time. For purely linear damping, with $f(t)\propto e^{-d_1t}$, we have $\beta(t)=d_1/\omega_0$. Therefore, in this case, condition \eqref{wdcond} implies that $d_1/\omega_0\ll1$. It is reasonable to expect that the addition of other types of damping to linear damping can only increase the value of $\beta(t)$. Consequently, one may expect that condition \eqref{wdcond} also implies $d_1/\omega_0\ll1$ when all three types of damping are present, as we demonstrate rigorously below. Thus, one simple way to obtain a reasonable $f(t)$, with all three types of damping present, is to consider the weak damping limit and use the time averaged energy dissipation rate, i.e. the relation \eqref{PowerAv}. 

In this case, $\omega\approx\omega_0$ holds with a relative error of the order $(d_1/\omega_0)^2$, and for the time-averaging interval we can take $\Delta T=T_0/2=\pi/\omega_0$, i.e. the half-period of the undamped oscillator. Furthermore, we assume $f(t)$ and $df(t)/dt$ vary slowly compared to the trigonometric part of the ansatz; consequently, we treat them as constant when averaging over $[t, t+T_0/2]$. Therefore, up to the first order in $d_1/\omega_0$ and $\beta(t)$, the moving average of our modeled energy \eqref{energija} is
\begin{equation}
\langle E(t)\rangle\approx\frac{m\omega_0^2A^2}{2}f^2(t)\left[\,1+\beta(t)\langle\sin(2\omega_0t+2\varphi)\rangle\,\right]=\frac{m\omega_0^2A^2}{2}f^2(t)\,,
\label{Everage}
\end{equation}
since $\langle\sin(2\omega_0t+2\varphi)\rangle=0$. Thus, to approximate the left hand side of relation \eqref{PowerAv} we can use
\begin{equation}
\frac{d\langle E(t)\rangle}{dt}\approx m\omega_0^2A^2f(t)\frac{df(t)}{dt}\,.
\label{lijeva}
\end{equation}
Using the identity $A_1\sin\theta+A_2\cos\theta=\sqrt{A_1^2+A_2^2}\,\sin\left(\theta+\arctan(A_2/A_1)\right)$, where $A_{1,2}>0$, the velocity \eqref{Vansatz1} can be written as
\begin{equation}
v(t)=-\omega Af(t)\sqrt{1+\left(\frac{\omega_0}{\omega}\beta(t)\right)^2}\,\sin\left(\omega t+\varphi+\arctan\left(\frac{\omega_0}{\omega}\beta(t)\right)\right)
\label{Vansatz11}
\end{equation}
Thus, up to the first order in $d_1/\omega_0$ and $\beta(t)$, the terms $\langle|v(t)|^n\rangle$ on the right hand side of \eqref{PowerAv} can be approximated by 
\begin{equation}
\langle|v(t)|^n\rangle\approx\left(\omega_0Af(t)\right)^n\langle|\sin\left(\omega_0t+\varphi+\arctan\beta(t)\right)|^n\rangle\,\,\,,\text{where}\,\,\,n=\lbrace1,2,3\rbrace\,.
\label{desna2}
\end{equation}
Using \eqref{lijeva} and \eqref{desna2} in \eqref{PowerAv} we get
\begin{equation}
\frac{df(t)}{dt}=-\frac{\mu g}{\omega_0A}\langle|\sin\theta(t)|\rangle-\frac{b}{m}f(t)\langle\sin^2\theta(t)\rangle-\frac{D\omega_0A}{m}f^2(t)\langle|\sin\theta(t)|^3\rangle\,,
\label{firstF2}
\end{equation}
where $\theta(t)=\omega_0t+\varphi+\arctan\beta(t)$. When evaluating $\langle|\sin\theta(t)|^n\rangle$ in \eqref{firstF2}, we assume that the term $\arctan\beta(t)$ remains approximately constant over the intervals $[t, t+T_0/2]$ (which holds exactly for purely linear damping). Since $\langle|\sin(\omega_0t+const)|\rangle=2/\pi$, $\langle\sin^2(\omega_0t+const)\rangle=1/2$, and $\langle|\sin(\omega_0t+const)|^3\rangle=4/(3\pi)$, we get
%
\begin{equation}
\frac{df(t)}{dt}=-\left(d_2Af^2(t)+d_1f(t)+\frac{d_0}{A}\right)\,,
\label{Fdiff}
\end{equation}
where
\begin{equation}
d_0=\frac{2\mu g}{\pi\omega_0}\,,\,d_1=\frac{b}{2m}\,,\,d_2=\frac{4D\omega_0}{3\pi m}\,,
\label{coeffC}
\end{equation}
as the final differential equation for the function $f(t)$. With $f(t)$ obeying \eqref{Fdiff}, the weak damping parameter \eqref{beta} becomes
\begin{equation}
\beta(t)=\frac{1}{\omega_0}\left(d_2Af(t)+d_1+\frac{d_0}{Af(t)}\right)\,,
\label{beta2}
\end{equation}
i.e. the sum of the three damping parameters evaluated with the current amplitude $Af(t)$. Since all terms are positive, $\beta(t)\geq d_1/\omega_0$, so the condition \eqref{wdcond} indeed implies $d_1/\omega_0\ll1$. Each type of damping is strong at a different stage of the motion. The contribution of linear damping is constant. The contribution of quadratic damping is largest at the beginning, when the amplitude is largest, and decreases as the amplitude decays. The contribution of Coulomb damping grows as the amplitude decreases and diverges as $f(t)\rightarrow0$, i.e. close to the halt of the motion. This anticipates the two regimes discussed below: for strong quadratic damping only the first half-period violates \eqref{wdcond}, while Coulomb damping limits the accuracy of the approximation in the last half-periods of the motion.

We are now in a position to write the final equations from which we determine the constants $A$ and $\varphi$. Combining \eqref{pocuvjet1} and \eqref{pocuvjet2}, along with \eqref{Fdiff}, we get
\begin{equation}
\left[v_0+\left(d_2A+d_1+\frac{d_0}{A}\right)x_0\right]^2=\omega^2\left(A^2-x_0^2\right)
\label{amplituda}
\end{equation}
as the equation from which we determine $A$, and
\begin{equation}
\varphi=\arctan{\left(-\frac{v_0+\left(d_2A+d_1+\frac{d_0}{A}\right)x_0}{\omega x_0}\right)}
\label{faza}
\end{equation}
as the equation from which we can calculate the initial phase (once we have determined $A$ from \eqref{amplituda}). The first order differential equation \eqref{Fdiff} can be solved by separation of variables and integration, i.e. we have to solve 
\begin{equation}
\int_{f(0)}^{f(t)}\left(d_2Af^2(t')+d_1f(t')+\frac{d_0}{A}\right)^{-1}df(t')=-\int_0^tdt'\,.
\label{Integral}
\end{equation}
Thus, $f(t)$ in our ansatz \eqref{Xansatz1} is obtained by solving \eqref{Integral}. We note that $\tilde{f}(t)$ in ansatz \eqref{Xansatz0} is obtained by solving the integral of the same form as \eqref{Integral}, but with $A_0$ in place of $A$ \cite{LelasPezer2}. Thus, once $f(t)$ is obtained, one easily obtains the corresponding $\tilde{f}(t)$ by replacing $A$ in $f(t)$ with $A_0$. Furthermore, for initial conditions $(x_0=0,v_0>0)$ equations \eqref{amplituda} and \eqref{faza} give $A=v_0/\omega$ and $\varphi=-\pi/2$, while for the undamped oscillator we get $A_0=v_0/\omega_0$ and $\varphi_0=-\pi/2$. Thus, for this particular type of initial conditions, the approximate analytical solutions \eqref{Xansatz0} and \eqref{Xansatz1} are the same for combinations of damping forces that do not include linear damping (since in that case $\omega=\omega_0$ holds), but even then the corresponding velocities \eqref{Vansatz0} and \eqref{Vansatz1}, as well as the energies \eqref{Eansatz0} and \eqref{energija}, are not the same. In the case of combinations of damping forces that include linear damping, solutions \eqref{Xansatz0} and \eqref{Xansatz1} are different for this type of initial conditions, due to $\omega\neq\omega_0$ in the solution \eqref{Xansatz1}.    

We note here that in the case of a harmonic oscillator with purely linear damping (i.e., when $d_1>0$ and $d_0=d_2=0$), our approximate solution \eqref{Xansatz1}, with initial amplitude given by \eqref{amplituda}, initial phase given by \eqref{faza}, and amplitude decay $f(t)$ obtained by solving \eqref{Integral}, becomes
\begin{equation}
x(t)=\sqrt{x_0^2+\left(\frac{v_0+d_1x_0}{\omega}\right)^2}\,e^{-d_1 t}\cos\left(\omega t+\arctan\left(-\frac{v_0+d_1x_0}{\omega x_0}\right)\right)\,,
\label{egazktnoX}
\end{equation}
i.e. equal to an exact solution of the harmonic oscillator with linear damping, e.g. see \cite{Lelas2023}. Thus, although we derive the differential equation \eqref{Fdiff} for $f(t)$ in the weak damping limit, our new approach yields an approximate solution that coincides with the exact solution, which is worth highlighting.  

For an easier comparison of approximate analytical solutions obtained by the new approach presented here with the solutions obtained by the approach used in \cite{LelasPezer2}, we will consider a block-spring system with the same parameters as in \cite{LelasPezer2}, i.e. we consider a block of mass $m=1\,kg$ attached to a spring of stiffness $k=30\,N/m$, and we use $g=9.81\,m/s^2$. The corresponding undamped angular frequency is $\omega_0=5.48\,s^{-1}$ and the period of the undamped system is $T_0=1.15\,s$. 
These values are in line with experiments performed on block-spring systems; see, e.g., \cite{Kamela}. Throughout the paper, as in \cite{LelasPezer2}, we consider initial conditions that correspond to undamped solutions with $A_0=0.2\,m$ and various values of $\varphi_0$, e.g., for $(x_0=A_0,v_0=0)$ we have $\varphi_0=0$, for $(x_0=0,v_0=\omega_0A_0)$ we have $\varphi_0=-\pi/2$, etc. 

For this choice of the parameters of block-spring system and initial conditions, solutions of the form \eqref{Xansatz0}, obtained in \cite{LelasPezer2}, describe well the dynamics governed by the equation \eqref{HOeq} for a range of values of the damping constants $\mu$, $b$ and $D$ that correspond to
\begin{equation}
0<\frac{d_0}{A_0}\lesssim0.03\omega_0\,,\,0<d_1\lesssim0.1\omega_0\,,\,0<d_2A_0\lesssim0.1\omega_0\,.
\label{co}
\end{equation}
Values of $d_0/A_0$, $d_1$, and $d_2A_0$ are suitable for quantifying the damping strength, since the weak damping limit is characterized by $d_0/A_0\ll\omega_0$, $d_1\ll\omega_0$, and $d_2A_0\ll\omega_0$ \cite{LelasPezer2}. In what follows, we will compare the approximate analytical solutions of the form \eqref{Xansatz1} with the solutions of the form \eqref{Xansatz0} and with the corresponding numerical solutions for various values of $\mu$, $b$ and $D$, i.e. for various values of $d_0/A_0$, $d_1$ and $d_2A_0$. 

Our main goal is to obtain approximate analytical solutions of the equation of motion \eqref{HOeq} that are more accurate and valid for a wider range of damping constants than the solutions recently presented in \cite{LelasPezer2}. Among the cases where only one of these three types of damping acts on a harmonic oscillator, the case with quadratic damping is the only one for which the corresponding equation of motion cannot be solved exactly by analytical means. Therefore, we first focus on the harmonic oscillator damped only by a force quadratic in velocity, and subsequently include the other two types of damping in our analysis.

\section{Damping quadratic in velocity}
\label{quadratic}

If $D>0$ and $\mu=b=0$ (i.e. $d_2>0$, $d_0=d_1=0$, and $\omega=\omega_0$), equation \eqref{amplituda} becomes a quadratic equation in $A$ and we easily get
\begin{equation}
A=\frac{x_0v_0d_2+\sqrt{x_0^2v_0^2d_2^2+\left(\omega_0^2-x_0^2d_2^2\right)\left(v_0^2+\omega_0^2x_0^2\right)}}{\omega_0^2-x_0^2d_2^2}\,,
\label{amplitudaKV}
\end{equation}
while equation \eqref{faza} simplifies to
\begin{equation}
\varphi=\arctan{\left(-\frac{v_0+x_0d_2A}{\omega_0x_0}\right)}\,.
\label{fazaKV}
\end{equation}
In this case, solving the integral \eqref{Integral} gives
\begin{equation}
f(t)=\frac{1}{1+d_2At}\,.
\label{fKV}
\end{equation}
Thus, in case of damping quadratic in velocity, our approximate solution \eqref{Xansatz1} becomes
\begin{equation}
x(t)=\left(\frac{A}{1+d_2At}\right)\cos(\omega_0t+\varphi)\,,
\label{xKV}
\end{equation}
where $A$ and $\varphi$ are given by \eqref{amplitudaKV} and \eqref{fazaKV}. In this case, the approach used in \cite{LelasPezer2} gives
\begin{equation}
\tilde{x}(t)=\left(\frac{A_0}{1+d_2A_0t}\right)\cos(\omega_0t+\varphi_0)\,,
\label{xKV0}
\end{equation}
where $A_0$ and $\varphi_0$ are defined in \eqref{A0phi0}. 

\subsection{Initial conditions $(x_0>0,v_0=0)$}
\label{quadratic1}

\begin{figure}[h!t!]
\begin{center}
\includegraphics[width=0.48\textwidth]{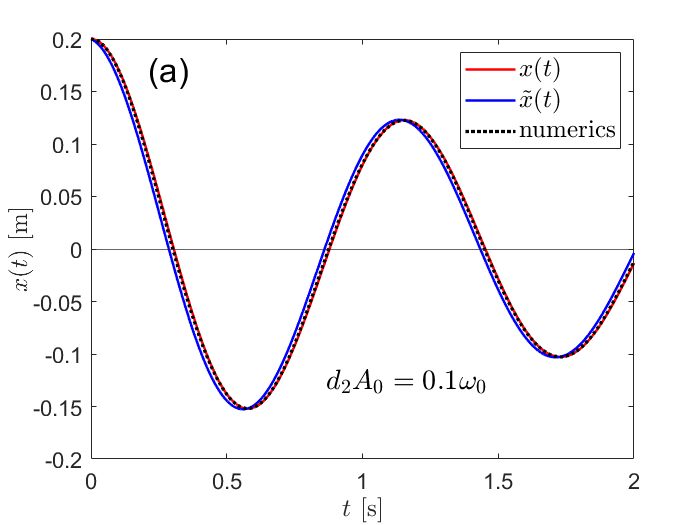}
\includegraphics[width=0.48\textwidth]{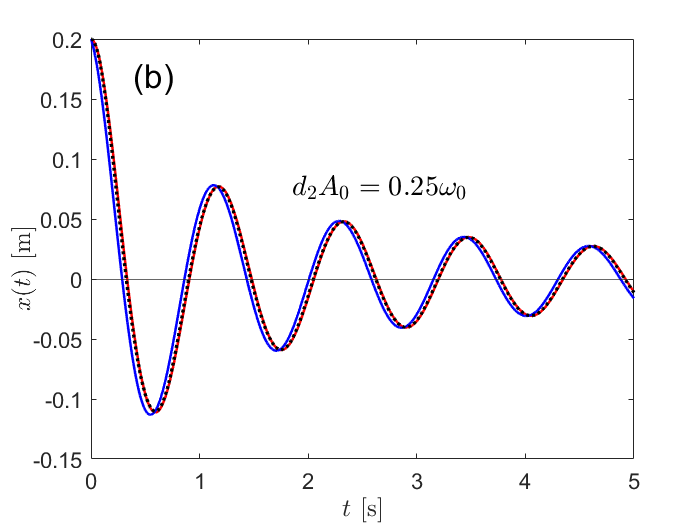}
\includegraphics[width=0.48\textwidth]{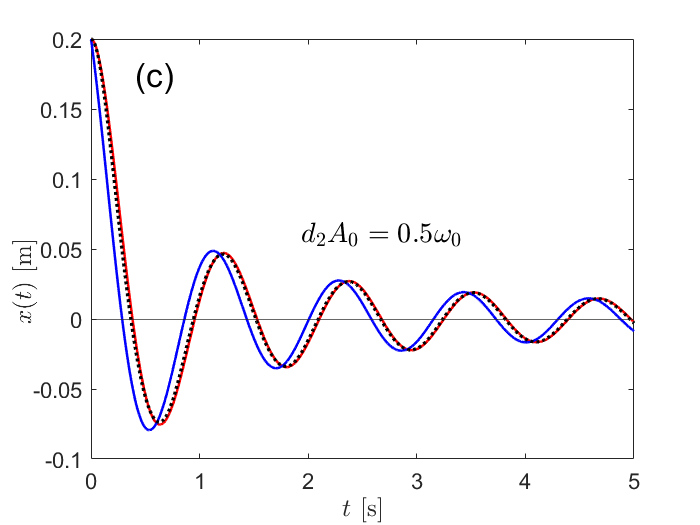}
\includegraphics[width=0.48\textwidth]{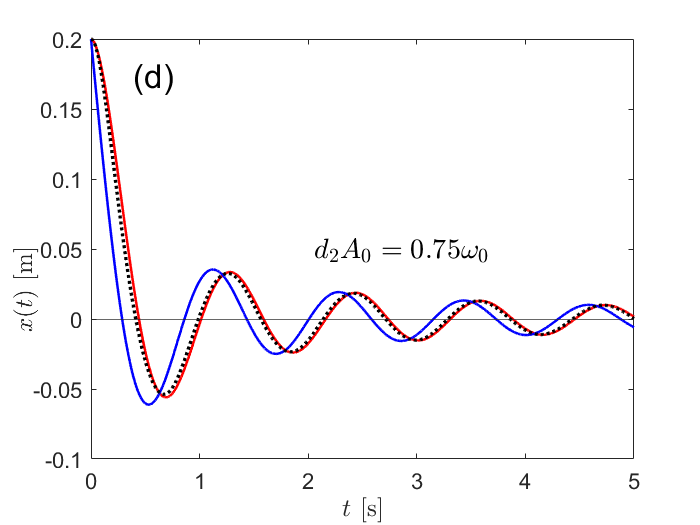}
\end{center}
\caption{Solid red curves show the solutions $x(t)$ with initial conditions $(x_0=0.2 m,v_0=0)$ and various values of the ratio $d_2 A_0/\omega_0$. Solid blue curves show the solutions $\tilde{x}(t)$ and black dotted curves show the numerical solutions for the same initial conditions and the corresponding values of the ratio $d_2 A_0/\omega_0$. See text for details.} 
\label{slika1}
\end{figure}

For this type of initial conditions \eqref{amplitudaKV} and \eqref{fazaKV} reduce to 
\begin{equation}
A=\frac{x_0}{\sqrt{1-\left(d_2x_0/\omega_0\right)^2}}\,\,\,\text{and}\,\,\,\varphi=\arctan{\left(-\frac{d_2A}{\omega_0}\right)}\,,
\label{AF1}
\end{equation}
while $A_0=x_0$ and $\varphi_0=0$. As noted earlier, throughout the paper we always choose initial conditions corresponding to $A_0=0.2\,m$, i.e. we take here $x_0=0.2\,m$. In Fig.\,\ref{slika1}(a)-(d) we show solutions $x(t)$ given by \eqref{xKV}, solutions $\tilde{x}(t)$ given by \eqref{xKV0} and the corresponding numerical solutions for $D/(kg/m)=\lbrace1.18,2.95,5.89,8.84\rbrace$, i.e. for $d_2A_0/\omega_0=\lbrace0.1,0.25,0.5,0.75\rbrace$, with initial conditions $(x_0=0.2m,v_0=0)$. We used the \emph{ode45} MATLAB function to obtain the corresponding numerical solutions of equation \eqref{HOeq}. We see that the solutions $x(t)$ agree better with the numerical results than $\tilde{x}(t)$ for all chosen values of the ratio $d_2A_0/\omega_0$. In Fig.\,\ref{slika1}(a) we have shown a shorter time span compared to Fig.\,\ref{slika1}(b)-(d), in order to make the difference between $x(t)$ and $\tilde{x}(t)$ more visible. 

\begin{figure}[h!t!]
\begin{center}
\includegraphics[width=0.48\textwidth]{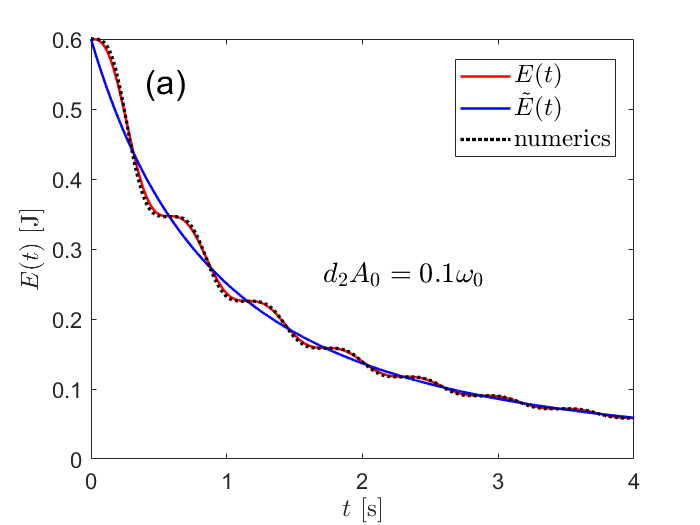}
\includegraphics[width=0.48\textwidth]{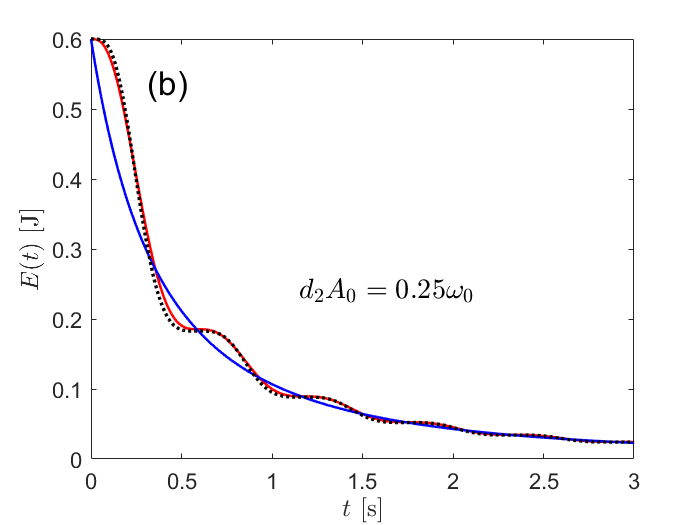}
\includegraphics[width=0.48\textwidth]{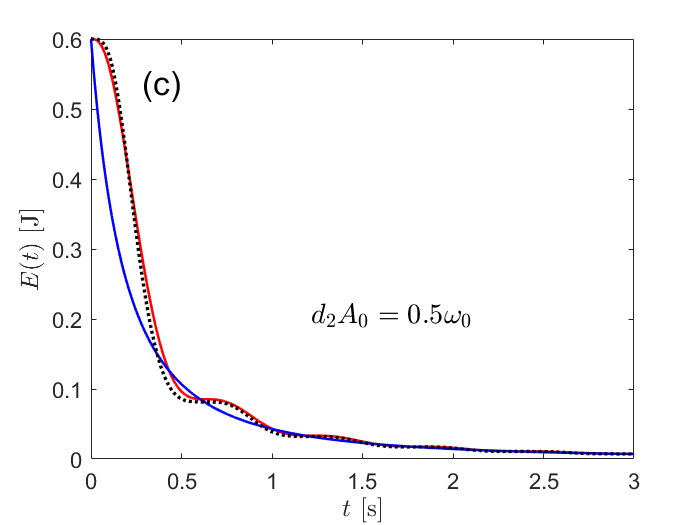}
\includegraphics[width=0.48\textwidth]{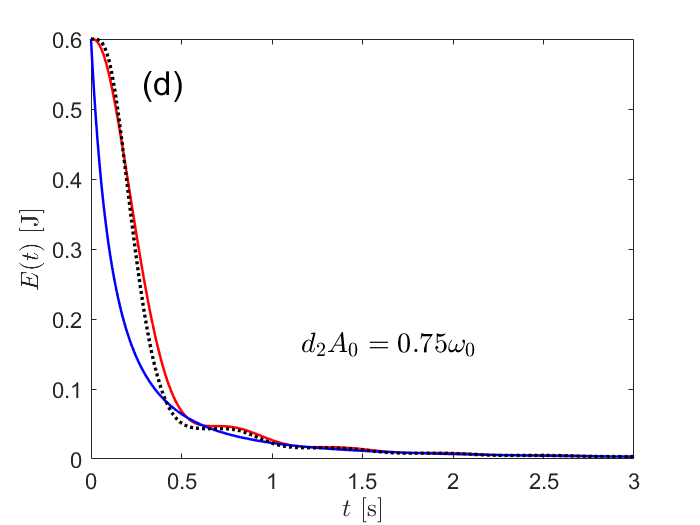}
\end{center}
\caption{Solid red curves show the energies $E(t)$ with initial conditions $(x_0=0.2 m,v_0=0)$ for the same values of the ratio $d_2 A_0/\omega_0$ as used in Fig.\,\ref{slika1}(a)-(d). Solid blue curves show the energies $\tilde{E}(t)$ and black dotted curves show the numerically obtained energies with the same initial conditions and for the corresponding values of the ratio $d_2 A_0/\omega_0$. See text for details.} 
\label{slika2}
\end{figure}

In Fig.\,\ref{slika2}(a)-(d) we show the energies $E(t)$ given by \eqref{energija}, the energies $\tilde{E}(t)$ given by \eqref{Eansatz0}, and the corresponding numerically obtained energies for $d_2A_0/\omega_0=\lbrace0.1,0.25,0.5,0.75\rbrace$ with initial conditions $(x_0=0.2m,v_0=0)$. For the chosen block-spring parameters and initial conditions, we have $E(0)=\tilde{E}(0)=0.6\,J$. 

\subsection{Initial conditions $(x_0=0,v_0>0)$}
\label{quadratic2}

\begin{figure}[h!t!]
\begin{center}
\includegraphics[width=0.48\textwidth]{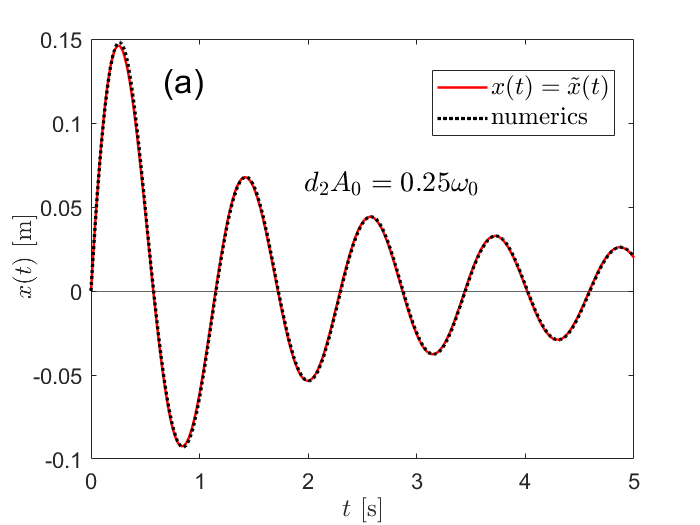}
\includegraphics[width=0.48\textwidth]{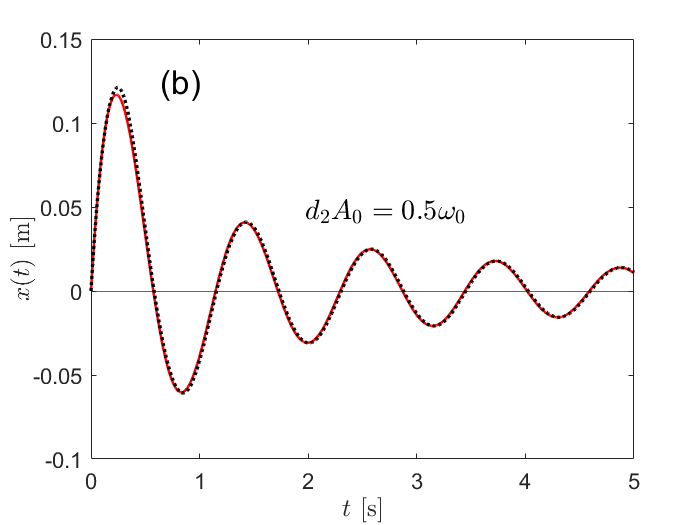}
\end{center}
\caption{Solid red curves show the solutions $x(t)$ with initial conditions $(x_0=0,v_0=1.10m/s)$ for two values of the ratio $d_2 A_0/\omega_0$. Dotted black curves show the numerical solutions for the same initial conditions and the corresponding values of the ratio $d_2 A_0/\omega_0$. See text for details.} 
\label{slika3}
\end{figure}

\begin{figure}[h!t!]
\begin{center}
\includegraphics[width=0.48\textwidth]{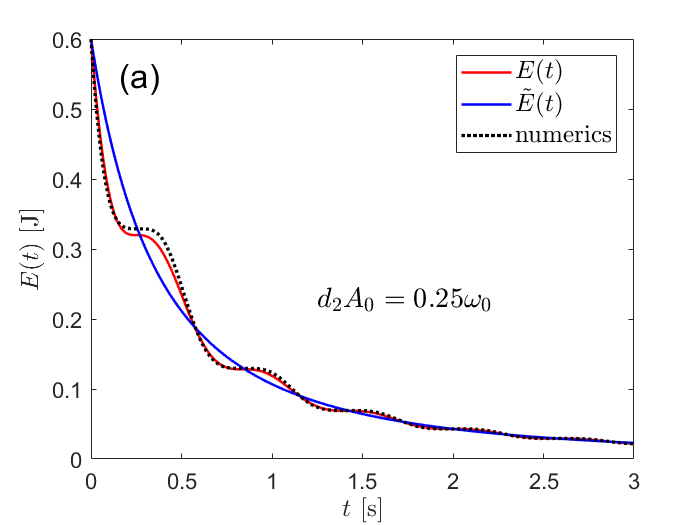}
\includegraphics[width=0.48\textwidth]{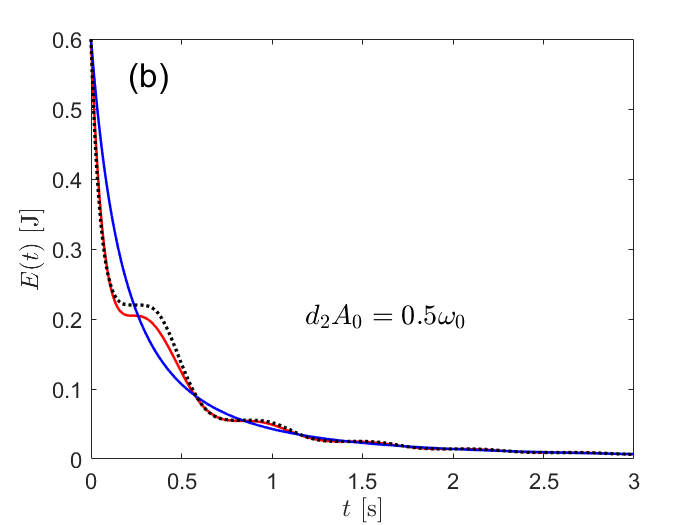}
\end{center}
\caption{Solid red curves show the energies $E(t)$ with initial conditions $(x_0=0,v_0=1.1m/s)$ for the same values of the ratio $d_2 A_0/\omega_0$ as used in Fig.\,\ref{slika3}(a) and (b). Solid blue curves show the energies $\tilde{E}(t)$ and black dotted curves show the numerically obtained energies with the same initial conditions and for the corresponding values of the ratio $d_2 A_0/\omega_0$. See text for details.} 
\label{slika4}
\end{figure}

For this type of initial conditions, \eqref{amplitudaKV} and \eqref{fazaKV} simplify to $A=A_0=v_0/\omega_0$ and $\varphi=\varphi_0=-\pi/2$, i.e. \eqref{xKV} and \eqref{xKV0} become
\begin{equation}
x(t)=\tilde{x}(t)=\left(\frac{A_0}{1+d_2A_0t}\right)\sin(\omega_0t)\,.
\label{xkvV0}
\end{equation}
We take here $v_0=\omega_0A_0=1.10m/s$. In Fig.\,\ref{slika3}(a) and (b) we show solutions \eqref{xkvV0} and the corresponding numerical solutions with initial conditions $(x_0=0,v_0=1.10m/s)$ for $d_2A_0/\omega_0=\lbrace0.25,0.5\rbrace$. In Fig.\,\ref{slika4}(a) and (b) we show the corresponding energies given by \eqref{energija} and \eqref{Eansatz0}, i.e. $E(t)$ and $\tilde{E}(t)$, and numerically obtained energies. Therefore, as we commented earlier, although the new approach and the approach used in \cite{LelasPezer2} give the same approximate analytical solutions in the case of quadratic damping with initial conditions $(x_0=0,v_0\neq0)$, the corresponding energies differ even in this case, i.e. the energy is in general more realistically modeled in the new approach.  

\subsection{Initial conditions $(x_0\neq0,v_0\neq0)$}
\label{quadratic3}

\begin{figure}[h!t!]
\begin{center}
\includegraphics[width=0.48\textwidth]{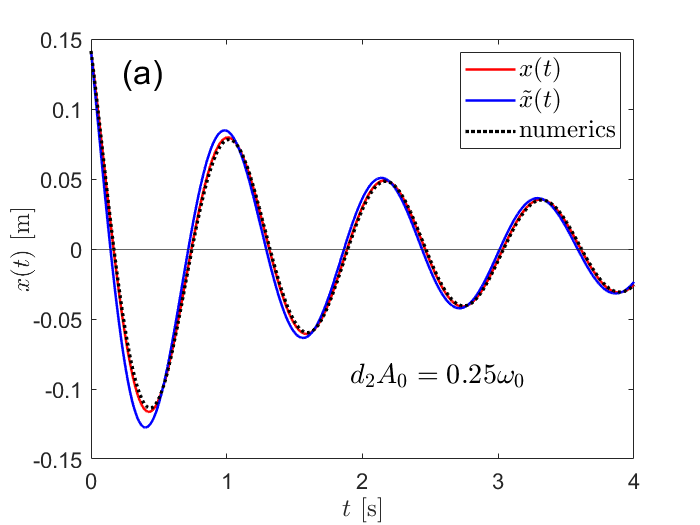}
\includegraphics[width=0.48\textwidth]{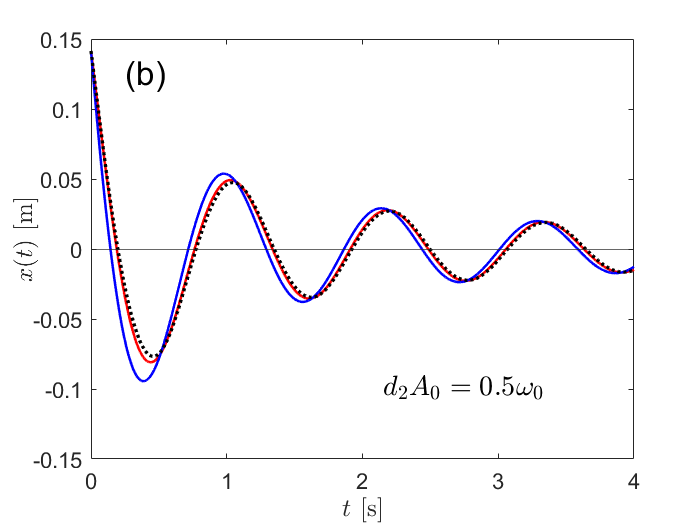}
\end{center}
\caption{Solid red curves show the solutions $x(t)$ with initial conditions $(x_0=0.14m,v_0=-0.77m/s)$ for two values of the ratio $d_2 A_0/\omega_0$. Solid blue curves show the solutions $\tilde{x}(t)$ and black dotted curves show the numerical solutions for the same initial conditions and the corresponding values of the ratio $d_2 A_0/\omega_0$. See text for details.} 
\label{slika5}
\end{figure}

\begin{figure}[h!t!]
\begin{center}
\includegraphics[width=0.48\textwidth]{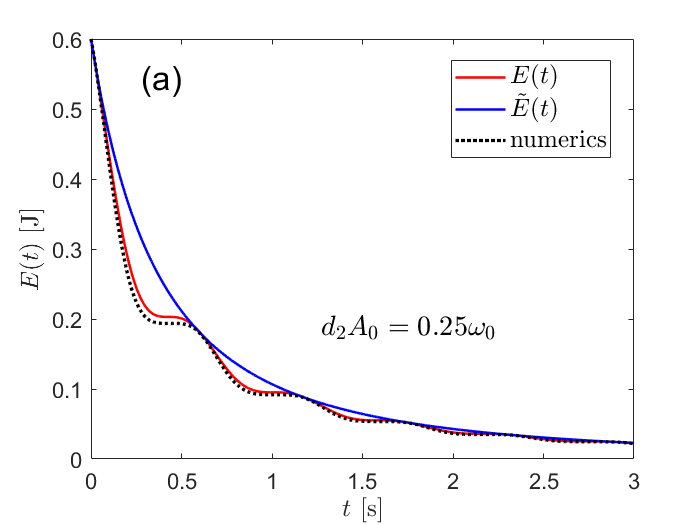}
\includegraphics[width=0.48\textwidth]{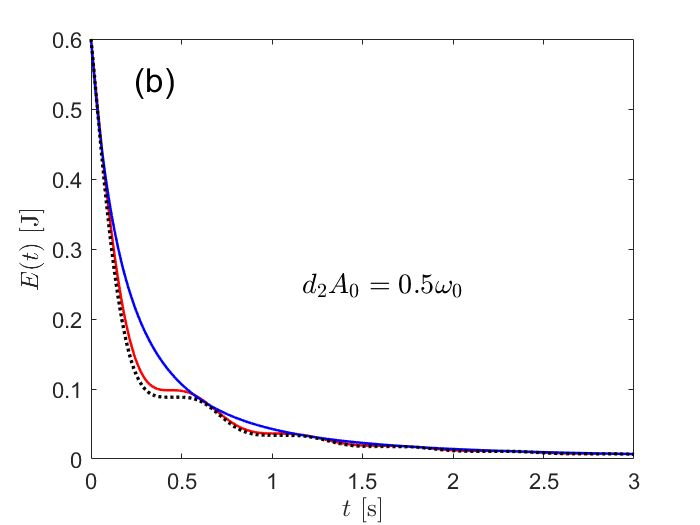}
\end{center}
\caption{Solid red curves show the energies $E(t)$ with initial conditions $(x_0=0.14,v_0=-0.77m/s)$ for the same values of the ratio $d_2 A_0/\omega_0$ as used in Fig.\,\ref{slika5}(a) and (b). Solid blue curves show the energies $\tilde{E}(t)$ and black dotted curves show the numerically obtained energies with the same initial conditions and for the corresponding values of the ratio $d_2 A_0/\omega_0$. See text for details.} 
\label{slika6}
\end{figure}

As an example, here we take $x_0=A_0\sqrt{2}/2=0.14m$ and $v_0=-\omega_0x_0=-0.77m/s$. For this choice, the initial potential energy equals the initial kinetic energy, and $\varphi_0=\pi/4$. In Fig.\,\ref{slika5}(a) and (b) we show solutions $x(t)$ given by \eqref{xKV}, solutions $\tilde{x}(t)$ given by \eqref{xKV0}, and the corresponding numerical solutions for $d_2A_0/\omega_0=\lbrace0.25,0.5\rbrace$ with chosen initial conditions. In Fig.\,\ref{slika6}(a) and (b) we show the corresponding energies given by \eqref{energija} and \eqref{Eansatz0}, i.e. $E(t)$ and $\tilde{E}(t)$, and the numerically obtained energies.

\subsection{Comments on results for harmonic oscillator with damping quadratic in velocity}
\label{quadratic4}

The results presented in this section show that the new approximate analytical solutions, i.e. solutions \eqref{xKV} with initial amplitude \eqref{amplitudaKV} and initial phase \eqref{fazaKV}, agree excellently with the numerical solutions in describing the free oscillations of a harmonic oscillator with damping quadratic in velocity for all types of initial conditions and for a range of values $0<D\lesssim5.89\,kg/m$ of the corresponding damping constant, i.e. for $0<d_2A_0/\omega_0\lesssim0.5$. For $d_2A_0/\omega_0>0.5$ the deviations of the solutions \eqref{xKV} from the numerical solutions become more pronounced, as the example in Fig.\,\ref{slika1}(d), with $D=8.84\,kg/m$ (i.e. $d_2A_0/\omega_0=0.75$), clearly shows.

Although the differential equation \eqref{Fdiff}, which gives us the function $f(t)$, was derived under the assumption of weak damping, i.e. under the weak damping condition $\beta(t)\ll1$ (see \eqref{beta}, we see that the solutions \eqref{xKV} describe the dynamics well for damping that we could characterize (at least) as moderately strong. For example, in \cite{CveticaninStrong} the elliptic harmonic balance method has been used to obtain approximate analytical solutions of the harmonic oscillator with strong quadratic damping, and the parameters used were such that the amplitude of the obtained approximate analytical solution, as well as the corresponding numerical solution, after the first period drops to somewhat less than $50\%$ of the initial value (see Fig. 6 in \cite{CveticaninStrong}). We can see, in Fig.\,\ref{slika1}(b), that for our choice of parameters, this happens for $d_2A_0/\omega_0=0.25$, and for $d_2A_0/\omega_0=0.5$ the amplitude after the first period drops to slightly less than $25\%$ of the initial value, as shown in Fig.\,\ref{slika1}(c). 

Furthermore, an ansatz of similar form as \eqref{Xansatz1} is used in \cite{CveticaninComp}, but with an unknown frequency (i.e. with $\omega\neq\omega_0$) and, by a more complex mathematical procedure, which also includes the time averaging over the trigonometric part of the ansatz, approximate solutions of the harmonic oscillator with damping quadratic in velocity were obtained with a frequency that is reduced in relation to the frequency of the undamped case. More precisely, the obtained frequency is $\omega=\omega_0-\Delta\omega_0$ \cite{CveticaninComp}, where $\Delta\omega_0>0$ is the term which depends on the constant of quadratic damping $D$ and on the magnitude of the initial amplitude. A brief numerical analysis of the free oscillations of the harmonic oscillator with damping quadratic in velocity presented in \cite{LelasPezer2} shows, for initial conditions $(x_0>0,v_0=0)$ and $0<d_2A_0/\omega_0\leq0.5$, that the quadratic damping has the greatest effect on the shift of the instants of the first zero crossing and the first turning point, while the time intervals between the subsequent zero crossings and the subsequent turning points remain approximately equal to the half-period of the undamped oscillator. Thus, the dynamics can be described well by retaining the frequency of the undamped system and accounting for a possible change in the initial phase, as we have done here. 

\begin{figure}[h!t!]
\begin{center}
\includegraphics[width=0.48\textwidth]{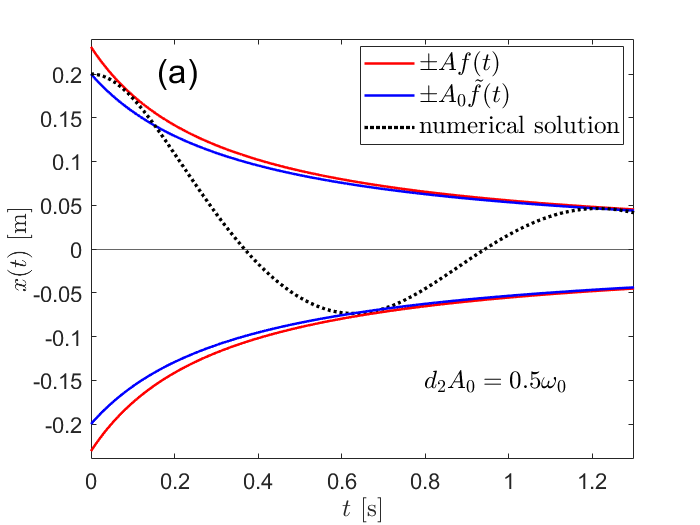}
\includegraphics[width=0.48\textwidth]{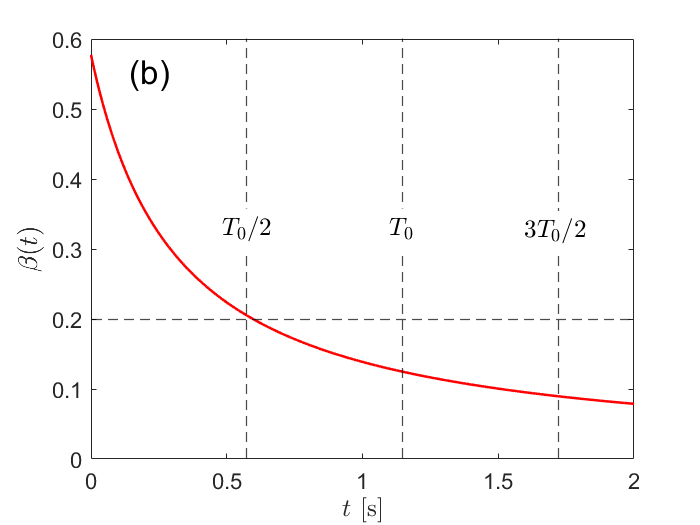}
\end{center}
\caption{(a) The envelopes of solutions $x(t)$ and $\tilde{x}(t)$ alongside the numerically obtained solution for $(x_0=0.2m, v_0=0)$ and $d_2A_0/\omega_0=0.5$. (b) The weak damping parameter $\beta(t)$, i.e. \eqref{beta3}, for $(x_0=0.2m, v_0=0)$ and $d_2A_0/\omega_0=0.5$. The black dashed vertical lines are positioned at $t=\lbrace T_0/2,\,T_0,\,3T_0/2\rbrace$, where $T_0=2\pi/\omega_0$. See text for details.} 
\label{slikaEnvelopa}
\end{figure}

We have already noted that, for purely linear damping, our new approach yields an exact analytical solution, even though we derived the amplitude decay function $f(t)$ under the weak damping assumption. For quadratic damping, no exact solution is available for comparison. Instead, in Fig.\,\ref{slikaEnvelopa}(a) we show the envelopes of solutions \eqref{xKV} and \eqref{xKV0} together with the numerical solution for $(x_0=0.2m, v_0=0)$ and $d_2A_0/\omega_0=0.5$. It can be seen that $Af(t)$, where $f(t)=\left(1+d_2A\,t\right)^{-1}$, indeed acts as an envelope for the numerical solution, while $A_0\tilde{f}(t)$, where $\tilde{f}(t)=\left(1+d_2A_0\,t\right)^{-1}$, fails to adequately capture the envelope-like behavior during the first half-period.
In Fig.\,\ref{slikaEnvelopa}(b) we show the weak damping parameter \eqref{beta}, which for quadratic damping, according to \eqref{beta2}, becomes
\begin{equation}
\beta(t)=\frac{d_2Af(t)}{\omega_0}=\frac{d_2A}{\omega_0\left(1+d_2A\,t\right)}\,.
\label{beta3}
\end{equation}
For $(x_0=0.2m, v_0=0)$ and $d_2A_0/\omega_0=0.5$, we have $A=0.231m$ and $d_2A/\omega_0=0.577$ in \eqref{beta3}. Our derivation of $f(t)$ is based on the weak damping condition $\beta(t)\ll1$. In Fig.\,\ref{slikaEnvelopa}(b), we can see that although the weak damping condition is not satisfied at the very beginning of the motion (since $\beta(0)=0.577$), $\beta(t)$ drops to about $0.2$ by the end of the first half-period and continues to decrease thereafter. Therefore, for strong quadratic damping, such as for $d_2A_0/\omega_0=0.5$, only the first half-period is problematic in this respect. However, our new solution $x(t)$ is constructed to match the initial conditions for any value of $\beta(0)$, so it aligns well with the numerical solution even during this initial stage. 

Capturing the dynamics during the first half-period is crucial for the overall accuracy, because quadratic damping is most pronounced precisely during that initial phase. After the first turning point the effect of quadratic damping is much weaker; even if the system is heavily damped initially, the amplitude at the first turning point drops to a level where the system operates in a moderate to weak damping regime. We therefore emphasize that the main novelty of our approach lies in the new initial amplitude $A\neq A_0$ and the new initial phase $\varphi\neq\varphi_0$, which make the approximate solution $x(t)$ satisfy both initial conditions for any $(x_0, v_0)$. This allows our solutions $x(t)$ to align excellently with the numerical solutions during the first half-period and, consequently, throughout the remaining dynamics even for rather strong damping.

\section{Combined linear and quadratic damping}
\label{LinearQuadratic}

If $b>0$, $D>0$ and $\mu=0$ (i.e. $d_1>0$, $d_2>0$, $d_0=0$ and $\omega=\sqrt{\omega_0^2-d_1^2}$), equation \eqref{amplituda} becomes a quadratic equation in $A$, i.e.
\begin{equation}
\left(\omega^2-d_2^2x_0^2\right)A^2-2\left(d_1d_2x_0^2+d_2x_0v_0\right)A-\left(d_1^2x_0^2+v_0^2+\omega^2x_0^2+2d_1x_0v_0\right)=0\,,
\label{amplitudaLINKV}
\end{equation}
which can be easily solved for any choice of initial conditions, and the initial phase \eqref{faza} becomes
\begin{equation}
\varphi=\arctan{\left(-\frac{v_0+x_0(d_1+d_2A)}{\omega x_0}\right)}\,.
\label{fazaLINKV}
\end{equation}
In this case, solving the integral \eqref{Integral} gives
\begin{equation}
f(t)=\frac{d_1}{e^{d_1t}\left(d_1+d_2A\right)-d_2A}\,.
\label{fLINKV}
\end{equation}
Thus, in the case of combined linear and quadratic damping, our approximate solution \eqref{Xansatz1} becomes
\begin{equation}
x(t)=\left(\frac{Ad_1}{e^{d_1t}\left(d_1+d_2A\right)-d_2A}\right)\cos(\omega t+\varphi)\,,
\label{xLINKV}
\end{equation}
where $A$ is obtained by solving \eqref{amplitudaLINKV} and $\varphi$ is given by \eqref{fazaLINKV}. In this case, the approach used in \cite{LelasPezer2} gives
\begin{equation}
\tilde{x}(t)=\left(\frac{A_0d_1}{e^{d_1t}\left(d_1+d_2A_0\right)-d_2A_0}\right)\cos(\omega_0t+\varphi_0)\,,
\label{xLINKV0}
\end{equation}
where $A_0$ and $\varphi_0$ are defined in \eqref{A0phi0}.

\subsection{Initial conditions $(x_0>0,v_0=0)$}
\label{linearquadratic1}

\begin{figure}[h!t!]
\begin{center}
\includegraphics[width=0.48\textwidth]{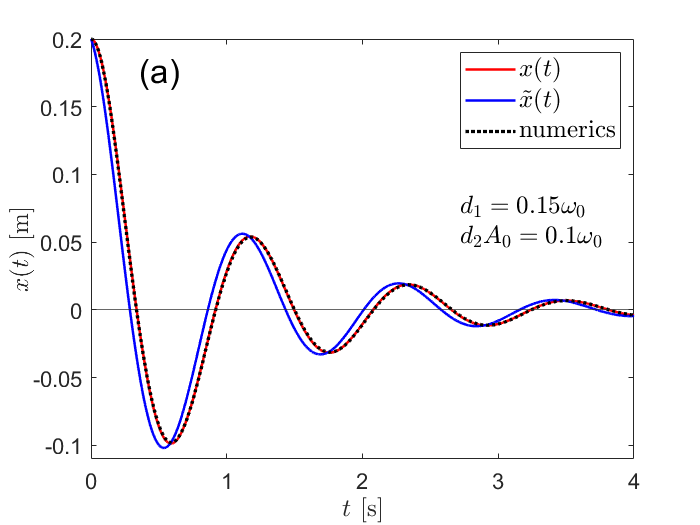}
\includegraphics[width=0.48\textwidth]{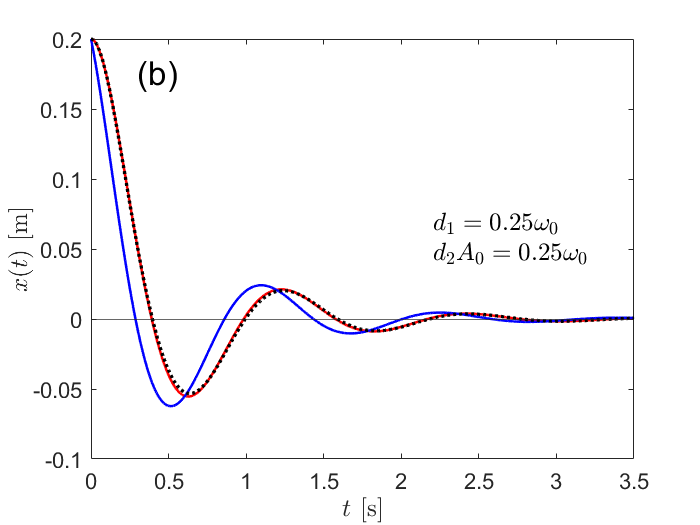}
\end{center}
\caption{Solid red curves show the solutions $x(t)$ with initial conditions $(x_0=0.2 m,v_0=0)$ for two combinations of $d_1/\omega_0$ and $d_2 A_0/\omega_0$. Solid blue curves show the solutions $\tilde{x}(t)$ and black dotted curves show the numerical solutions for the same initial conditions and the corresponding values of $d_1/\omega_0$ and $d_2A_0/\omega_0$. See text for details.} 
\label{slikaXLINKV1}
\end{figure}

\begin{figure}[h!t!]
\begin{center}
\includegraphics[width=0.48\textwidth]{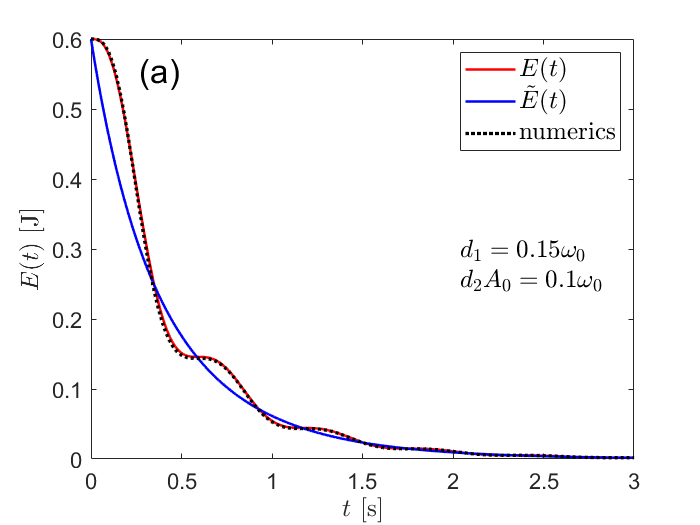}
\includegraphics[width=0.48\textwidth]{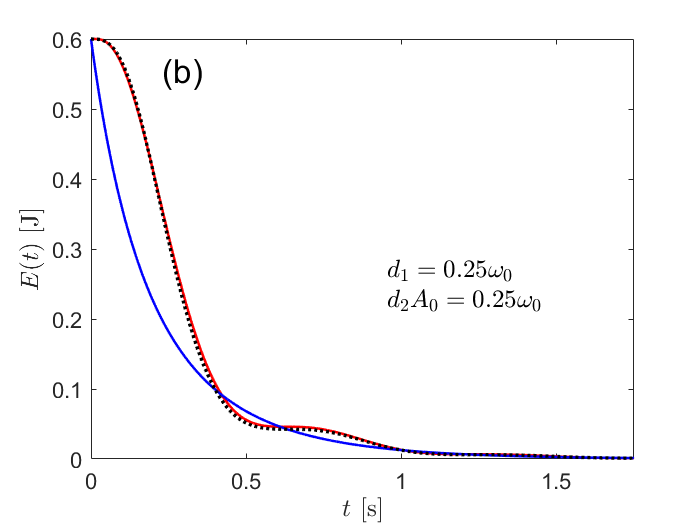}
\end{center}
\caption{Solid red curves show the energies $E(t)$ with initial conditions $(x_0=0.2 m,v_0=0)$ for the same values of the $d_1/\omega_0$ and $d_2A_0/\omega_0$ as used in Fig.\,\ref{slikaXLINKV1}(a) and (b). Solid blue curves show the energies $\tilde{E}(t)$, and black dotted curves show the numerically obtained energies for the corresponding initial conditions and damping values. See text for details.} 
\label{slikaELINKV1}
\end{figure}

For this type of initial conditions, the solution of the quadratic equation \eqref{amplitudaLINKV} and initial phase \eqref{fazaKV} are  
\begin{equation}
A=\frac{d_1d_2x_0^2+\omega x_0\sqrt{\omega^2+d_1^2-d_2^2x_0^2}}{\omega^2-d_2^2x_0^2}\,\,\,\text{and}\,\,\,\varphi=\arctan{\left(-\frac{d_1+d_2A}{\omega}\right)}\,,
\label{AF1LINKV}
\end{equation}
while $A_0=x_0$ and $\varphi_0=0$. In Fig.\,\ref{slikaXLINKV1}(a) and (b) we show solutions $x(t)$ given by \eqref{xLINKV}, solutions $\tilde{x}(t)$ given by \eqref{xLINKV0} and the corresponding numerical solutions for combined damping with $d_1/\omega_0=\lbrace0.15,0.25\rbrace$ and $d_2A_0/\omega_0=\lbrace0.1,0.25\rbrace$, for initial conditions $(x_0=0.2m,v_0=0)$. We can see that the solutions $x(t)$ overlap perfectly with the numerical solutions. 
In Fig.\,\ref{slikaELINKV1}(a) and (b) we show the energies $E(t)$ given by \eqref{energija}, the energies $\tilde{E}(t)$ given by \eqref{Eansatz0}, and the numerically obtained energies corresponding to the solutions shown in Fig.\,\ref{slikaXLINKV1}(a) and (b). We can see that the energies $E(t)$ overlap excellently with the numerically obtained energies. 

\begin{figure}[h!t!]
\begin{center}
\includegraphics[width=0.48\textwidth]{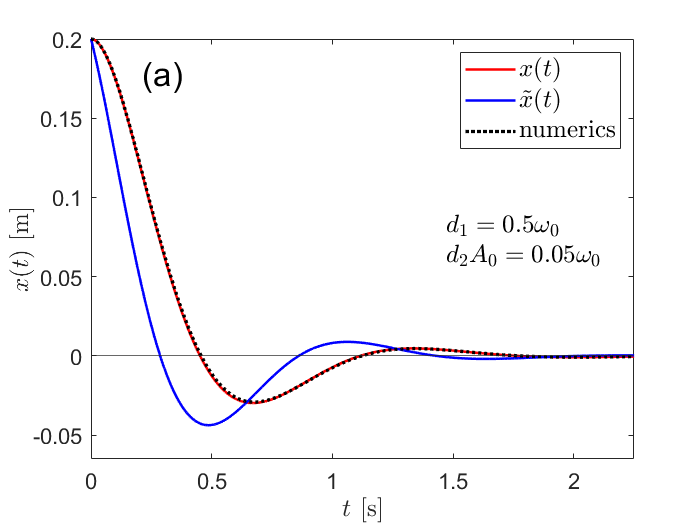}
\includegraphics[width=0.48\textwidth]{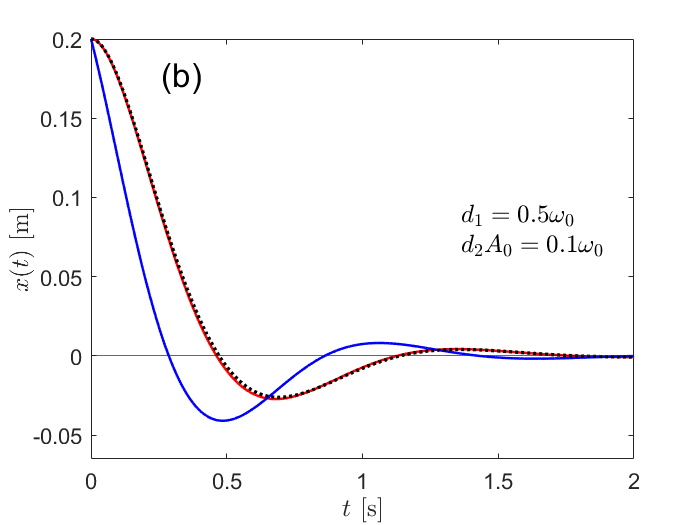}
\end{center}
\caption{Solid red curves show the solutions $x(t)$ with initial conditions $(x_0=0.2 m,v_0=0)$ for combinations with rather high value of $d_1/\omega_0$ and relatively weak values of $d_2 A_0/\omega_0$. Solid blue curves show the solutions $\tilde{x}(t)$ and black dotted curves show the numerical solutions for the same initial conditions and the corresponding values of $d_1/\omega_0$ and $d_2A_0/\omega_0$. See text for details.} 
\label{strongLINKV}
\end{figure}

We note here that, in the limit $d_2\rightarrow0$, our solution $x(t)$, given by \eqref{xLINKV}, becomes the exact solution for a harmonic oscillator with linear damping only. Therefore, when considering combinations with relatively weak quadratic damping, the linear damping can be strong while $x(t)$ still provides an excellent description of the dynamics. As an example, in Fig.\,\ref{strongLINKV}(a) and (b) we show solutions $x(t)$, solutions $\tilde{x}(t)$ and the corresponding numerical solutions for combined damping with $d_1/\omega_0=0.5$ and $d_2A_0/\omega_0=\lbrace0.05,0.1\rbrace$, for initial conditions $(x_0=0.2m,v_0=0)$. We can see in Fig.\,\ref{strongLINKV}(a) that the solution $x(t)$ with $d_1/\omega_0=0.5$ and $d_2A_0/\omega_0=0.05$ overlaps perfectly with the numerical solutions, while in Fig.\,\ref{strongLINKV}(b) we can notice a slight mismatch between $x(t)$ and the numerical solution in case with $d_1/\omega_0=0.5$ and $d_2A_0/\omega_0=0.1$, which can be attributed to stronger quadratic damping compared to the value used in Fig.\,\ref{strongLINKV}(a).     

\subsection{Initial conditions $(x_0=0,v_0>0)$}
\label{linearquadratic2}

\begin{figure}[h!t!]
\begin{center}
\includegraphics[width=0.48\textwidth]{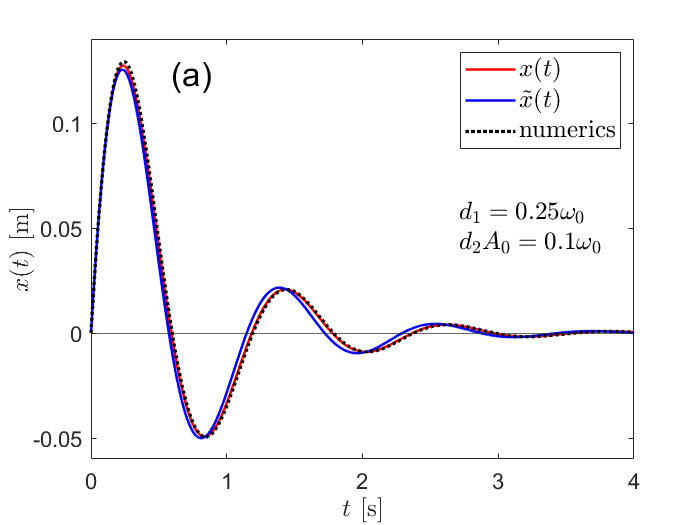}
\includegraphics[width=0.48\textwidth]{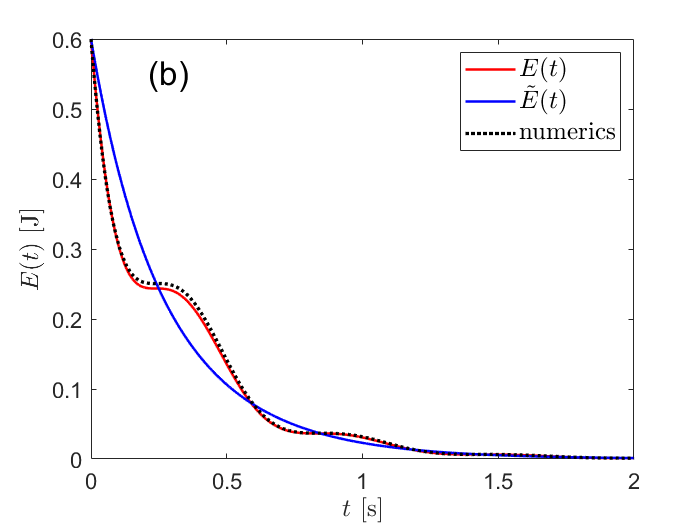}
\end{center}
\caption{(a) Solid red curve shows the solution $x(t)$ with initial conditions $(x_0=0,v_0=1.10m/s)$ for $d_1/\omega_0=0.25$ and $d_2A_0/\omega_0=0.1$. The solid blue curve shows the corresponding $\tilde{x}(t)$ solution, and the dotted black curve shows the numerical solution.(b) The solid red curve shows the energy $E(t)$, the solid blue curve shows the energy $\tilde{E}(t)$, and the dotted black curve shows the numerically obtained energy corresponding to the solutions shown in figure (a). See text for details.} 
\label{xV0LINKV}
\end{figure}

For this type of initial conditions, the initial amplitudes and the initial phases simplify to $A=v_0/\omega$, $A_0=v_0/\omega_0$, and $\varphi=\varphi_0=-\pi/2$, i.e. \eqref{xLINKV} becomes
\begin{equation}
x(t)=\frac{v_0}{\omega}\left(\frac{d_1}{e^{d_1t}\left(d_1+d_2v_0/\omega\right)-d_2v_0/\omega}\right)\sin(\omega t)\,,
\label{xlinkvV0}
\end{equation}
while $\tilde{x}(t)$, i.e. solution \eqref{xLINKV0}, becomes the same as \eqref{xlinkvV0} but with $\omega_0$ in place of $\omega$. We take $v_0=1.10m/s$. In Fig.\,\ref{xV0LINKV}(a) we show solutions $x(t)$, $\tilde{x}(t)$ and the corresponding numerical solution with initial conditions $(x_0=0,v_0=1.10m/s)$ for $d_1/\omega_0=0.25$ and $d_2A_0/\omega_0=0.1$. In Fig.\,\ref{xV0LINKV}(b) we show the corresponding energies given by \eqref{energija} and \eqref{Eansatz0}, i.e. $E(t)$ and $\tilde{E}(t)$, and numerically obtained energies. 

\subsection{Initial conditions $(x_0\neq0,v_0\neq0)$}
\label{linearquadratic3}

\begin{figure}[h!t!]
\begin{center}
\includegraphics[width=0.48\textwidth]{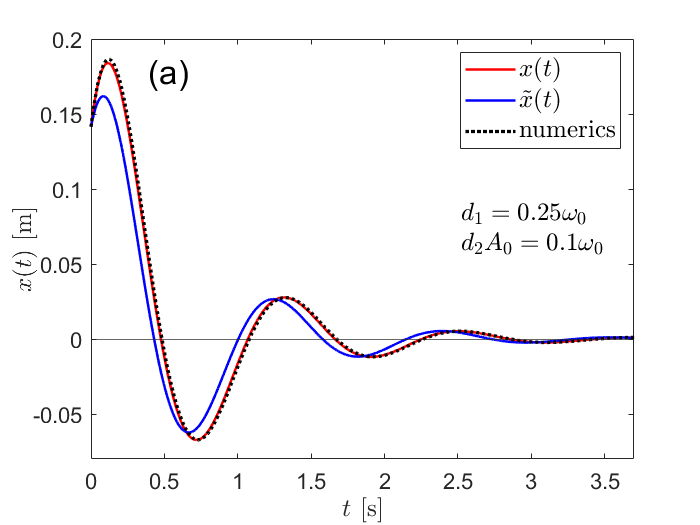}
\includegraphics[width=0.48\textwidth]{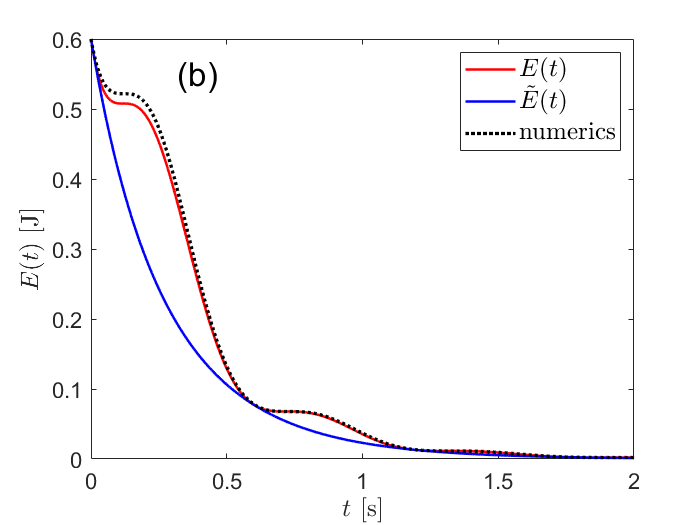}
\end{center}
\caption{(a) Solid red curve shows the solution $x(t)$ with initial conditions $(x_0=0.14m,v_0=0.77m/s)$ for $d_1/\omega_0=0.25$ and $d_2A_0/\omega_0=0.1$. The solid blue curve shows the corresponding $\tilde{x}(t)$ solution, and the dotted black curve shows the numerical solution.(b) The solid red curve shows the energy $E(t)$, the solid blue curve shows the energy $\tilde{E}(t)$, and the dotted black curve shows the numerically obtained energy corresponding to the solutions shown in figure (a). See text for details.} 
\label{slikaLINKVx0v0}
\end{figure}

For this type of initial conditions, solving the quadratic equation \eqref{amplitudaLINKV} gives
\begin{equation}
A=\frac{d_1d_2x_0^2+d_2x_0v_0+\sqrt{\left(d_1d_2x_0^2+d_2x_0v_0\right)^2+\left(\omega^2-d_2^2x_0^2\right)\left(v_0^2+\omega^2x_0^2+2d_1x_0v_0+d_1^2x_0^2\right)}}{\omega^2-d_2^2x_0^2}\,
\label{AlinkvadX0V0}
\end{equation}
for the initial amplitude, while the initial phase is given by \eqref{fazaLINKV}. For example, we take $x_0=A_0\sqrt{2}/2=0.14m$ and $v_0=\omega_0x_0=0.77m/s$. For this choice of initial conditions, the initial potential energy equals the initial kinetic energy, and $\varphi_0=-\pi/4$. In Fig.\,\ref{slikaLINKVx0v0}(a) we show solution $x(t)$ given by \eqref{xLINKV}, solution $\tilde{x}(t)$ given by \eqref{xLINKV0}, and the corresponding numerical solution with initial conditions $(x_0=0.14m,v_0=0.77m/s)$ for $d_1/\omega_0=0.25$ and $d_2A_0/\omega_0=0.1$. In Fig.\,\ref{slikaLINKVx0v0}(b) we show the energies given by \eqref{energija} and \eqref{Eansatz0}, i.e. $E(t)$ and $\tilde{E}(t)$, and numerically obtained energies that correspond to the solutions shown in Fig.\,\ref{slikaLINKVx0v0}(a). 

\subsection{Comments on the results for a harmonic oscillator with combined linear and quadratic damping}
\label{linearquadratic4}

The results presented in this section show that the new approximate analytical solutions $x(t)$, i.e. solutions \eqref{xLINKV}, with initial amplitude obtained by solving the quadratic equation \eqref{amplitudaLINKV} and initial phase given by \eqref{fazaLINKV}, agree excellently with the numerical solutions in describing the free oscillations of a harmonic oscillator with combined linear and quadratic damping for a range of values of the damping constant $b$ and $D$ corresponding to $0<d_1/\omega_0\lesssim0.25$ and $0<d_2A_0/\omega_0\lesssim0.25$. Furthermore, for sufficiently small quadratic damping, e.g., with $0<d_2A_0/\omega_0\lesssim0.1$, our solution $x(t)$ describes the dynamics very accurately, even with strong linear damping ($d_1/\omega_0\lesssim0.5$). As mentioned previously, in the limit $d_2\rightarrow0$, $x(t)$ reduces to the exact solution for a harmonic oscillator with linear damping only. Thus, $x(t)$ can remain accurate even for higher values of $d_1/\omega_0$, provided that the quadratic damping remains sufficiently weak.


\section{Combination of all three types of damping}
\label{AllThree}

If $\mu>0$, $b>0$ and $D>0$ (i.e. $d_0>0$, $d_1>0$ and $d_2>0$), the equation for the initial amplitude \eqref{amplituda} becomes a quartic equation
\begin{equation}
\left(\frac{\omega^2}{x_0^2}-d_2^2\right)A^4-2d_2\left(\frac{v_0}{x_0}+d_1\right)A^3-\left(\omega^2+\left(\frac{v_0}{x_0}+d_1\right)^2+2d_0d_2\right)A^2-2d_0\left(\frac{v_0}{x_0}+d_1\right)A-d_0^2=0\,,
\label{amplitudaLQC}
\end{equation}
and the initial phase is given by the expression \eqref{faza}, which we repeat here for the reader's convenience
\begin{equation}
\varphi=\arctan{\left(-\frac{v_0+\left(d_0/A+d_1+d_2A\right)x_0}{\omega\,x_0}\right)}\,.
\label{fazaLQC}
\end{equation}
Depending on the value of $C=4d_2d_0-d_1^2$, the solutions of the integral \eqref{Integral} have three different forms. For $C>0$, we get 
\begin{equation}
f(t)=\frac{1}{2d_2A}\left[\sqrt{C}\tan\left(-\frac{\sqrt{C}}{2}t+\arctan\frac{2d_2A+d_1}{\sqrt{C}}\right)-d_1\right]\,,
\label{rjes1}
\end{equation}
for $C<0$  
\begin{equation}
f(t)=\frac{1}{2d_2A}\left[\sqrt{|C|}\coth\left(\frac{\sqrt{|C|}}{2}t+\text{arccoth}\frac{2d_2A+d_1}{\sqrt{|C|}}\right)-d_1\right]\,,
\label{rjes2}
\end{equation}
and for $C=0$ ($d_1=2\sqrt{d_2d_0}\,$)  
\begin{equation}
f(t)=\frac{1}{\sqrt{d_2A}}\left[\left(\sqrt{d_2A}\,t+\frac{1}{\sqrt{d_2A}+\sqrt{d_0/A}}\right)^{-1}-\sqrt{d_0/A}\right]\,.
\label{rjes3}
\end{equation}
The three forms \eqref{rjes1}--\eqref{rjes3} can be written as a single expression derived from \eqref{rjes1}. Using the addition formula for the tangent and identity $C+d_1(d_1+2d_2A)=2d_2(2d_0+d_1A)$, we turn \eqref{rjes1} into  
\begin{equation}
f(t)=\frac{\cos\left(\Omega\,t\right)-p\,\sin\left(\Omega\,t\right)/\Omega}{\cos\left(\Omega\,t\right)+q\,\sin\left(\Omega\,t\right)/\Omega}\,,
\label{rjes11}
\end{equation}
where we introduced the notation $\Omega=\sqrt{C}/2$, $p=\left(d_0/A+d_1/2\right)$, and $q=\left(d_1/2+d_2A\right)$. For $C<0$ (i.e. for $\Omega=i|\Omega|$, where $i=\sqrt{-1}$) the relations $\cos\left(\Omega\,t\right)=\cosh\left(|\Omega|\,t\right)$ and $\sin\left(\Omega\,t\right)/\Omega=\sinh\left(|\Omega|\,t\right)/|\Omega|$ hold, and we can use them to show that the expressions \eqref{rjes11} and \eqref{rjes2} are equivalent in this case. To obtain the expression \eqref{rjes3}, valid for $C=0$, we can take the limit $\Omega\rightarrow0$ of the expression \eqref{rjes11} and obtain $f(t)=(1-p\,t)/(1+q\,t)$, which, after some algebraic manipulation (and using $d_1=2\sqrt{d_2d_0}\,$), can be shown to be equivalent to the expression \eqref{rjes3}. 

Thus, we can directly take \eqref{rjes11} as the expression for $f(t)$ in all cases with $C\neq0$. The specific case where $C=0$ requires precisely tuned values for $d_0$, $d_1$, and $d_2$, making it impossible in practice to strictly satisfy this equality; however, in theoretical research, we can easily choose the parameter values to achieve this equality exactly, in which case, as previously explained, the solution is again contained in expression \eqref{rjes11}. In that sense, $f(t)$ given by expression \eqref{rjes11} is valid for all values of $C$.   


As is well known \cite{Lapidus, AviAJP, Grk2, Coulomb} the dynamics of a harmonic oscillator damped (only) by Coulomb damping halts in finite time and, in general, at some $x_{halt}\neq0$, for which $|k\,x_{halt}|\leq\mu mg$ holds. We take here, for simplicity, that the dynamic and static friction coefficients are the same, i.e. both are equal to $\mu$. Of course, when we add the other two types of damping to Coulomb damping, the dynamics of free vibrations also halts in finite time and generally at a finite displacement, except that in this case it is not possible to calculate the exact instant and position of the halt analytically (due to quadratic damping in the equation of motion). 
The function $f(t)$, given by \eqref{rjes11}, monotonically decreases from value $f(0)=1$ to value $f(\tau)=0$, where 
\begin{equation}
\tau=\frac{2}{\sqrt{C}}\,\arctan\left(\frac{A\sqrt{C}}{2d_0+d_1A}\right)\,.
\label{tau1}
\end{equation}
For $t>\tau$, the function $f(t)$ becomes increasingly negative, which corresponds to an increase in the amplitude of oscillations and is not physical. In this case, we will take $t=\tau$ as an approximation of the instant when the dynamics halts. 

We note that the expression \eqref{tau1} is valid for both positive and negative values of $C$. Although, for $C<0$ (i.e. when $4d_2d_0<d_1^2$), the argument of the inverse tangent function becomes imaginary, the value of the expression \eqref{tau1} is real. This follows from the identity $$\frac{\arctan(i\alpha)}{i\sqrt{|C|}}=\frac{\text{arctanh}(\alpha)}{\sqrt{|C|}}\,,$$ where $\alpha=A\sqrt{|C|}/(2d_0+d_1A)$. Since $A$, $d_0$, $d_1$, and $d_2$ are all positive, we have $0<\alpha<1$ whenever $C<0$, ensuring that $\text{arctanh}(\alpha)$ remains well-defined and real. Furthermore, the limit $C\rightarrow0$ of the expression \eqref{tau1} exists and is given by $\tau=2A/(2d_0+d_1A)$, providing a continuous extension of \eqref{tau1} to $C=0$. Thus, the expression for $\tau$ given by \eqref{tau1} can be used for all values of $C$, with the value at $C=0$ defined by continuity. Therefore, our approximate solution \eqref{Xansatz1} can be written as
%
%
\begin{equation}
\begin{split}
x(t)&=Af(t)\cos(\omega t+\varphi)\,\Theta(\tau-t)\\
&\quad=A\left(\frac{\cos\left(\Omega\,t\right)-p\,\sin\left(\Omega\,t\right)/\Omega}{\cos\left(\Omega\,t\right)+q\,\sin\left(\Omega\,t\right)/\Omega}\right)\cos(\omega t+\varphi)\Theta(\tau-t)\,,
\end{split}
\label{xLQC}
\end{equation}
where 
\begin{equation} \label{theta}
    \Theta(\tau-t) = \begin{cases}
\begin{tabular}{@{}cl@{}}
   $1$\, & if\, $t \leq\tau$ \\
    $0$\, & if\, $t>\tau$\,.
\end{tabular}
    \end{cases}
    \end{equation}
The step function $\Theta(\tau-t)$ is simply added by hand to exclude the non-physical part of the solution. The velocity corresponding to \eqref{xLQC} is
\begin{equation}
v(t)=\left(-\omega Af(t)\sin(\omega t+\varphi)+A\frac{df(t)}{dt}\cos(\omega t+\varphi)\right)\Theta(\tau-t)\,,
\label{vLQC}
\end{equation}
where $f(t)$ is given by \eqref{rjes11}, and the corresponding energy $E(t)=kx(t)^2/2+mv(t)^2/2$ is calculated with \eqref{xLQC} and \eqref{vLQC}. In this case, the approach used in \cite{LelasPezer2} gives $\tilde{f}(t)$ of the same form as \eqref{rjes11}, but with $A_0$ in place of $A$, i.e. the approximate solution \eqref{Xansatz0} becomes
\begin{equation}
\tilde{x}(t)=A_0\left(\frac{\cos\left(\Omega\,t\right)-\tilde{p}\,\sin\left(\Omega\,t\right)/\Omega}{\cos\left(\Omega\,t\right)+\tilde{q}\,\sin\left(\Omega\,t\right)/\Omega}\right)\cos(\omega_0 t+\varphi_0)\Theta(\tilde{\tau}-t)\,,
\label{xLQC0}
\end{equation}
where $\tilde{p}=\left(d_0/A_0+d_1/2\right)$ and $\tilde{q}=\left(d_1/2+d_2A_0\right)$, while $\Theta(\tilde{\tau}-t)$ is of the same form as $\Theta(\tau-t)$ but with
\begin{equation}
\tilde{\tau}=\frac{2}{\sqrt{C}}\,\arctan\left(\frac{A_0\sqrt{C}}{2d_0+d_1A_0}\right)\,
\label{tau0}
\end{equation}
in place of $\tau$. The energy corresponding to solution \eqref{xLQC0} is $\tilde{E}(t)=\left(m\omega_0^2A_0^2\tilde{f}(t)^2/2\right)\Theta(\tilde{\tau}-t)$. 

We note that, in what follows, instead of finding analytical solutions of the quartic equation for the initial amplitude, i.e. equation \eqref{amplitudaLQC}, we solve it numerically. For example, if we consider the case with $(x_0=0.2m, v_0=0)$, $d_0/A_0=0.03\omega_0$, $d_1=0.15\omega_0$, and $d_2A_0=0.15\omega_0$, equation \eqref{amplitudaLQC} becomes $$716.2500\,A^4-6.7500\,A^3-30.2700\,A^2-0.0540\,A-0.0011=0\,,$$ where we did not write the physical units and the resulting $A$ is in meters. Using \emph{roots} MATLAB function, we get four solutions $$A/m=\lbrace0.2113, -0.2001, -0.0009+0.0059i, -0.0009-0.0059i\rbrace\,.$$ We expected one of the solutions to be close to $A_0=0.2m$, therefore we take the value $A=0.2113m$ as our initial amplitude and the corresponding initial phase \eqref{fazaLQC} is $\varphi=-0.3284$. 

Of course, there are ways to find an approximate analytical solution of a quartic equation \eqref{amplitudaLQC}. For example, we see that $(A-A_0)/A_0\ll1$ is valid, so we could write the initial amplitude as $A=A_0(1+\varepsilon)$, where $\varepsilon\ll1$, insert that expression into the quartic equation and then use $A^n\approx A_0^n(1+n\varepsilon)$ where $n=\lbrace1,2,3,4\rbrace$, i.e. ignore terms of order $\varepsilon^2$ and higher. This gives a linear equation in $\varepsilon$, and $A_0(1+\varepsilon)$ is a solid analytical approximation to the solution of the quartic equation \eqref{amplitudaLQC}. We could go a step further and retain the terms of order $\varepsilon^2$, yielding an even more precise solution while keeping the procedure tractable. Still, for simplicity, we used the numerical approach with the \emph{roots} MATLAB function.


\subsection{Initial conditions $(x_0>0,v_0=0)$}
\label{Clinearquadratic1}

\begin{figure}[h!t!]
\begin{center}
\includegraphics[width=0.48\textwidth]{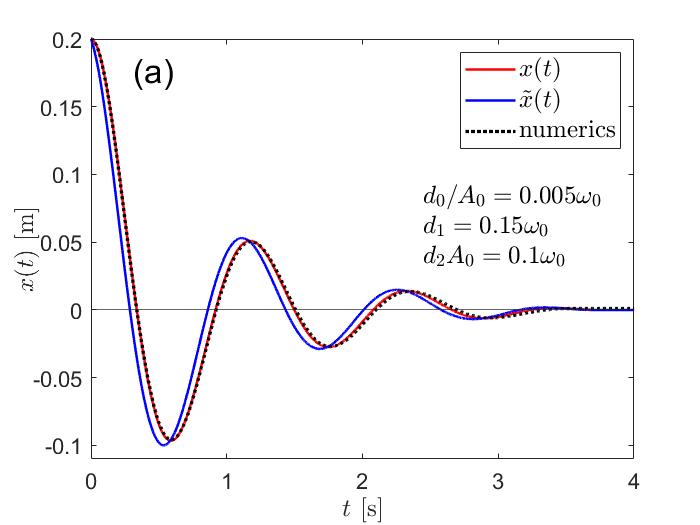}
\includegraphics[width=0.48\textwidth]{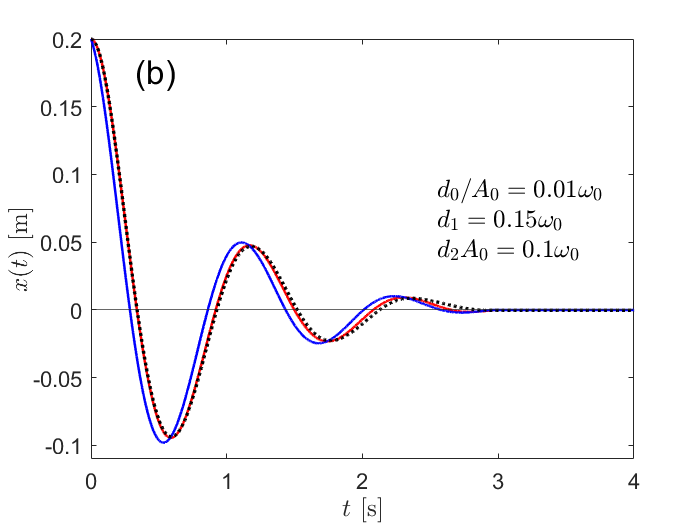}
\includegraphics[width=0.48\textwidth]{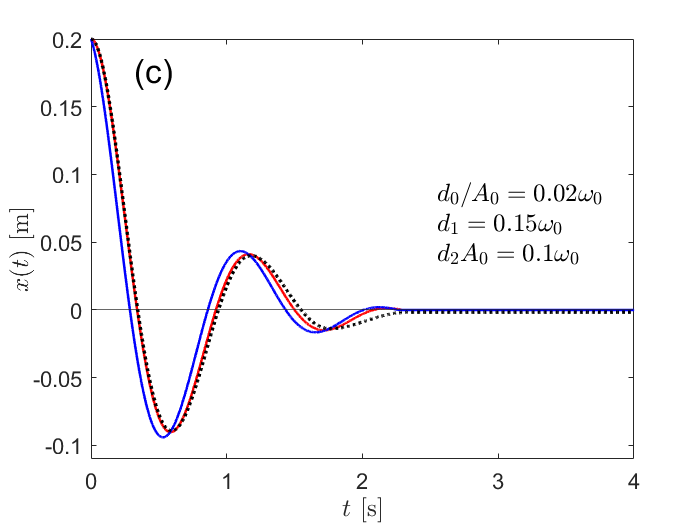}
\includegraphics[width=0.48\textwidth]{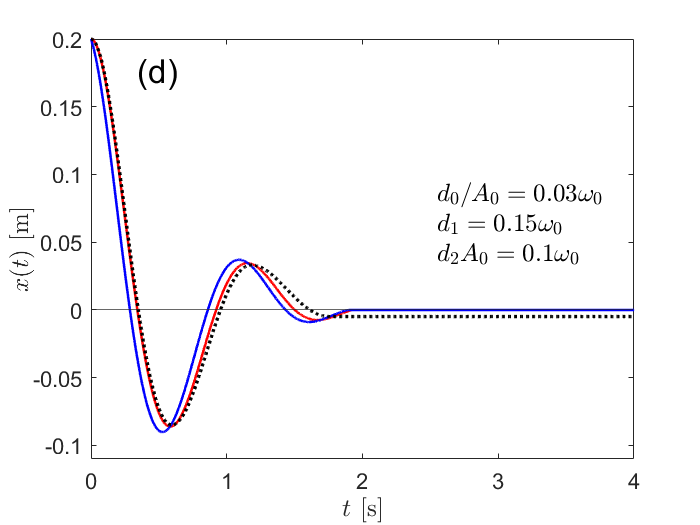}
\end{center}
\caption{Solid red curves show the solutions $x(t)$, i.e. \eqref{xLQC}, with initial conditions $(x_0=0.2m, v_0=0)$ for $d_0/(\omega_0A_0)=\lbrace0.005, 0.01, 0.02, 0.03\rbrace$, $d_1/\omega_0=0.15$ and $d_2A_0/\omega_0=0.1$. Solid blue curves show the solutions $\tilde{x}(t)$, i.e. \eqref{xLQC0}, and black dotted curves show the numerical solutions for the same initial conditions and the corresponding values of $d_0/(\omega_0A_0)$, $d_1/\omega_0$ and $d_2A_0/\omega_0$. See text for details.} 
\label{slikaCLQ1}
\end{figure}

In Fig.\,\ref{slikaCLQ1}(a)--(d) we show solutions $x(t)$ given by \eqref{xLQC}, solutions $\tilde{x}(t)$ given by \eqref{xLQC0} and the corresponding numerical solutions for combined damping with $d_0/(\omega_0A_0)=\lbrace0.005, 0.01, 0.02, 0.03\rbrace$, $d_1/\omega_0=0.15$ and $d_2A_0/\omega_0=0.1$, for initial conditions $(x_0=0.2m,v_0=0)$. Here, we used the same values for the linear and quadratic damping, i.e. the same $d_1$ and $d_2A_0$, and the same time span as in Fig.\,\ref{slikaXLINKV1}(a), in order to clearly demonstrate the effect of adding Coulomb damping on the dynamics of the system. In Fig.\,\ref{slikaXLINKV1}(a) oscillations are visible over the entire displayed time span (i.e. for $0\leq t\leq 4s$). In Fig.\,\ref{slikaCLQ1}(a) we can see that already adding Coulomb damping with $d_0/(\omega_0A_0)=0.005$ (which corresponds to $\mu=0.0048$ for the block-spring system we consider here) clearly halts the oscillations at about $t\approx3.5s$. In Fig.\,\ref{slikaCLQ1}(b)-(d), we see that as the value of $d_0/(\omega_0A_0)$ increases, the system comes to a halt sooner and it also becomes evident that it can halt at $x_{halt}\neq0$. 
Although our approximate analytical solutions necessarily come to rest at the equilibrium position $x(\tau)=0$, meaning that they cannot accurately model $x_{halt}\neq0$, we see that both $\tau$ and $\tilde{\tau}$, given by \eqref{tau1} and \eqref{tau0}, provide good approximations of the halt instant. Since $|x_{halt}|\leq\mu mg/k$, the error of the approximate solutions \eqref{xLQC} and \eqref{xLQC0} in the halting position compared to the numerical solutions can be at most $\Delta x=\pm\mu mg/k$, which amounts, e.g., to $\Delta x=\pm0.016\,x_0$ in case $d_0/(\omega_0A_0)=0.01$ or to $\Delta x=\pm0.047\,x_0$ in case $d_0/(\omega_0A_0)=0.03$.

\begin{figure}[h!t!]
\begin{center}
\includegraphics[width=0.48\textwidth]{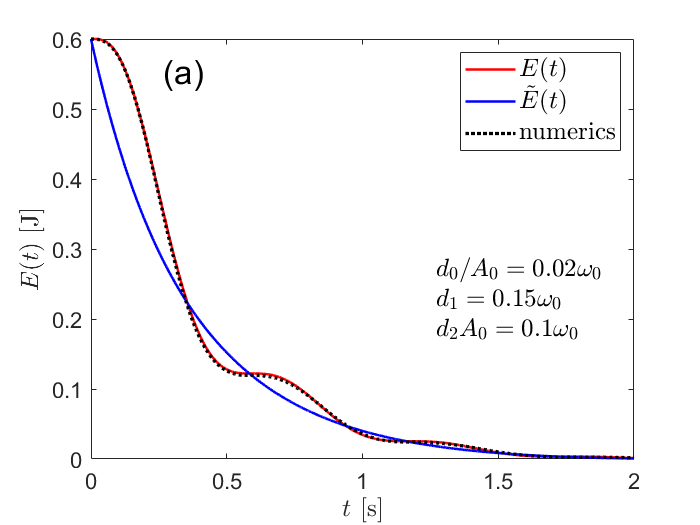}
\includegraphics[width=0.48\textwidth]{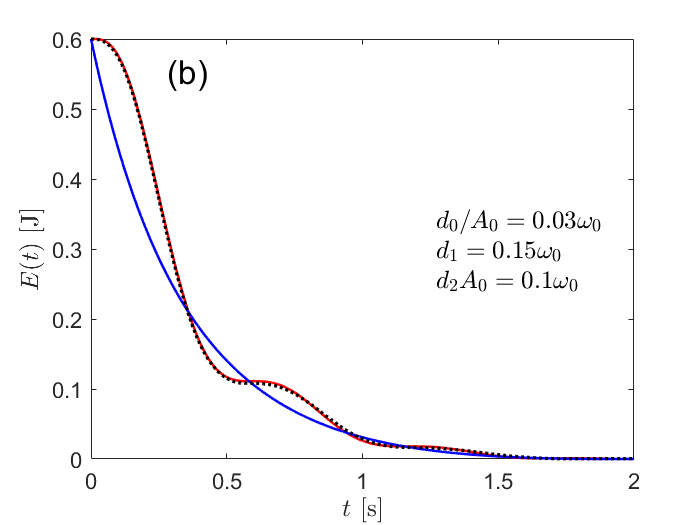}
\end{center}
\caption{Energies corresponding to solutions shown in Fig.\,\ref{slikaCLQ1}(c) and (d). See text for details.} 
\label{slikaCLQ2}
\end{figure}

In \cite{LelasPezer2}, by analyzing the exact solution of the harmonic oscillator with Coulomb damping only, it was shown that Coulomb damping does not change the frequency of the harmonic oscillator, but changes the instants of zero crossings, i.e. each subsequent zero crossing has an increasing time shift in relation to the zero crossing instants of the undamped oscillator, and therefore this time shift is greatest for the last zero crossing. An ansatz of form \eqref{Xansatz0} or \eqref{Xansatz1} cannot encapsulate this effect of Coulomb damping \cite{LelasPezer2}, since in these cases the interval between two consecutive zero crossings is fixed, i.e. always equal to a half-period. Thus, in addition to the error in the halting position, our approximate solutions \eqref{xLQC} also have errors in zero crossing instants compared to numerical solutions. For example, in Fig.\,\ref{slikaCLQ1}(a) and (b), we can see that the first two zero crossings of $x(t)$ overlap excellently with zero crossings of the numerical solutions, while in Fig.\,\ref{slikaCLQ1}(c) and (d) this is not the case, and the reason for this is the larger shift in zero crossings due to stronger Coulomb damping compared to the values used in Fig.\,\ref{slikaCLQ1}(a) and (b). In Fig.\,\ref{slikaCLQ2}(a) and (b) we show energies corresponding to solutions shown in Fig.\,\ref{slikaCLQ1}(c) and (d). We see that the region where our approximate solutions $x(t)$ do not overlap well with the numerical solutions (i.e. for $t\gtrsim1.25s$ in Fig.\,\ref{slikaCLQ1}(c) and (d)) corresponds to energies that are a small or negligible fraction of the initial energy.  

\begin{figure}[h!t!]
\begin{center}
\includegraphics[width=0.48\textwidth]{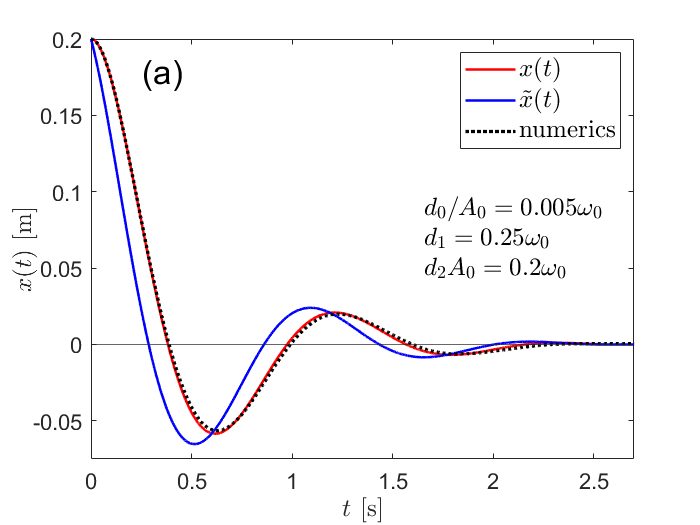}
\includegraphics[width=0.48\textwidth]{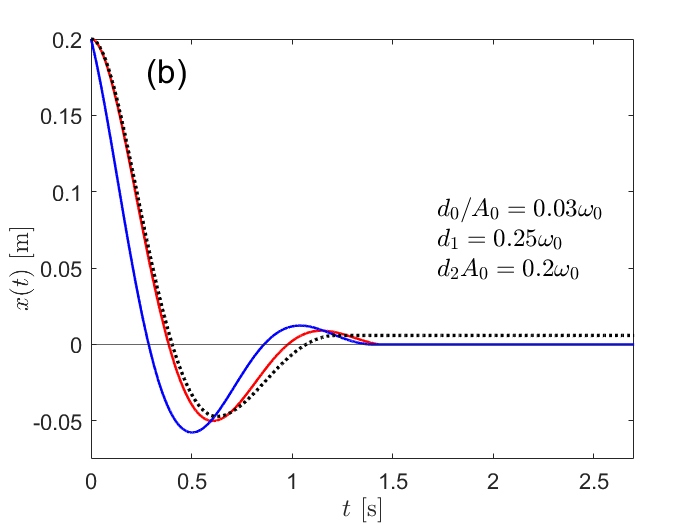}
\includegraphics[width=0.48\textwidth]{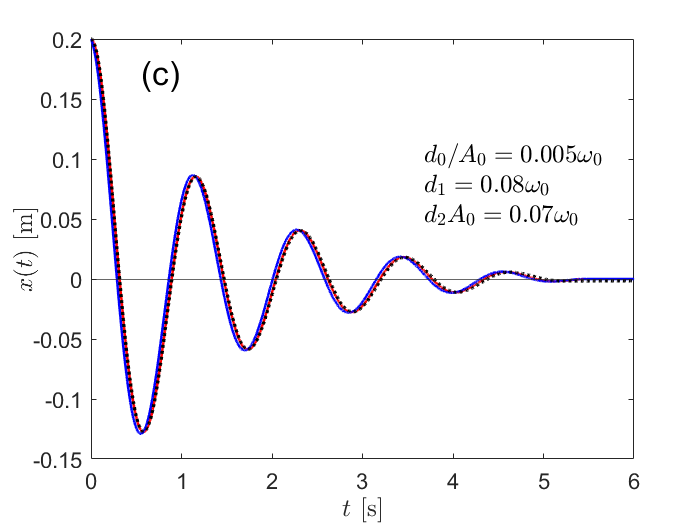}
\includegraphics[width=0.48\textwidth]{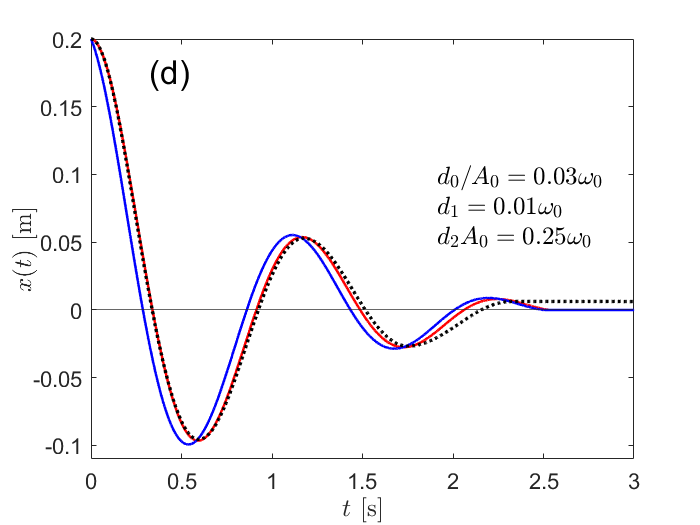}
\end{center}
\caption{Solid red curves show the solutions $x(t)$ with initial conditions $(x_0=0.2m, v_0=0)$ for various values of $d_0/(\omega_0A_0)$, $d_1/\omega_0$ and $d_2A_0/\omega_0$. Solid blue curves and black dotted curves show the corresponding $\tilde{x}(t)$ and numerical solutions. See text for details.} 
\label{slikaCLQdodana}
\end{figure}

Before moving on to other types of initial conditions, in Fig.\,\ref{slikaCLQdodana}(a)--(d) we show more examples with various values of damping. These, along with the examples presented previously, show that approximate analytical solutions $x(t)$ are valid for a much wider range of $d_1$ and $d_2A_0$ than solutions $\tilde{x}(t)$. However, for Coulomb damping, both types of solutions are considerably limited. We take that $x(t)$ describes the dynamics with Coulomb damping reasonably well for $0<d_0/(\omega_0A_0)\lesssim0.03$, just as was estimated for $\tilde{x}(t)$ in \cite{LelasPezer2}.

\subsection{Initial conditions $(x_0=0,v_0>0)$}
\label{Clinearquadratic2}

\begin{figure}[h!t!]
\begin{center}
\includegraphics[width=0.48\textwidth]{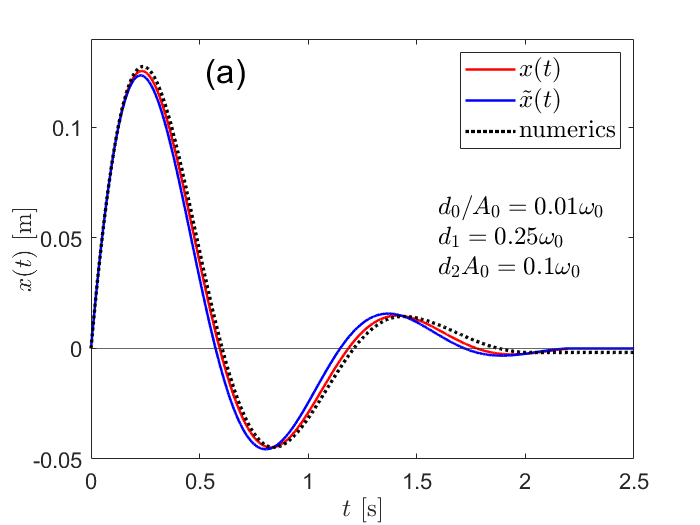}
\includegraphics[width=0.48\textwidth]{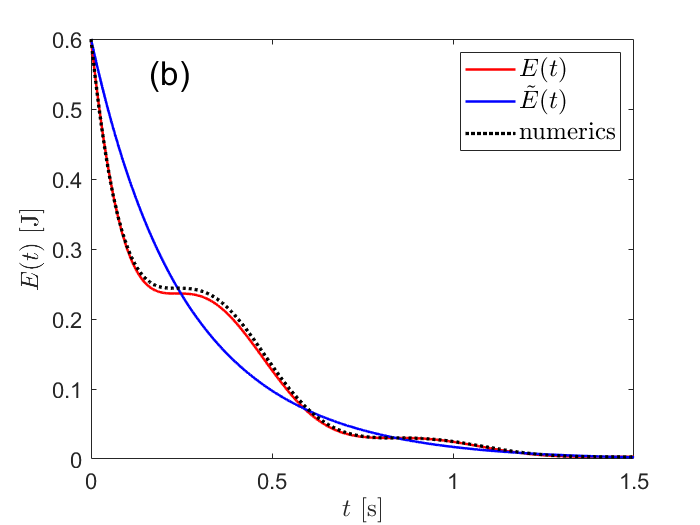}
\end{center}
\caption{(a) Solid red curve shows the solution $x(t)$ with initial conditions $(x_0=0, v_0=1.10m/s)$ for $d_0/(\omega_0A_0)=0.01$, $d_1/\omega_0=0.25$ and $d_2A_0/\omega_0=0.1$. The solid blue curve and the black dotted curve show the corresponding $\tilde{x}(t)$ solution and the numerical solution. (b) Energies corresponding to solutions shown in figure (a). See text for details.} 
\label{slikaCLQ3}
\end{figure}

For this type of initial conditions, the initial amplitudes and initial phases are simply $A=v_0/\omega$, $A_0=v_0/\omega_0$, and $\varphi=\varphi_0=-\pi/2$. Thus, in this case \eqref{xLQC} becomes
\begin{equation}
x(t)=\frac{v_0}{\omega}\left(\frac{\cos\left(\Omega\,t\right)-p\,\sin\left(\Omega\,t\right)/\Omega}{\cos\left(\Omega\,t\right)+q\,\sin\left(\Omega\,t\right)/\Omega}\right)\sin(\omega t)\Theta(\tau-t)\,,
\label{xClinkvV0}
\end{equation}
while $\tilde{x}(t)$, i.e. solution \eqref{xLQC0}, obtains the same form as \eqref{xClinkvV0} but with $\omega_0$ in place of (each) $\omega$, and with the corresponding values of $\tilde{p}$, $\tilde{q}$ and $\tilde{\tau}$. In Fig.\,\ref{slikaCLQ3}(a) we show solutions $x(t)$, $\tilde{x}(t)$ and the corresponding numerical solutions with initial conditions $(x_0=0,v_0=1.10m/s)$ for $d_0/(\omega_0A_0)=0.01$, $d_1/\omega_0=0.25$ and $d_2A_0/\omega_0=0.1$. Again, we can see a strong influence of Coulomb damping on the shift of zero crossings of the numerically obtained solutions (towards the end of the dynamics) compared to our solutions. In Fig.\,\ref{slikaCLQ3}(b) we show the energies $E(t)$, $\tilde{E}(t)$ and the numerically obtained energies corresponding to the solutions shown in Fig.\,\ref{slikaCLQ3}(a). 
%
%
%

\subsection{Initial conditions $(x_0\neq0,v_0\neq0)$}
\label{Clinearquadratic3}

\begin{figure}[h!t!]
\begin{center}
\includegraphics[width=0.48\textwidth]{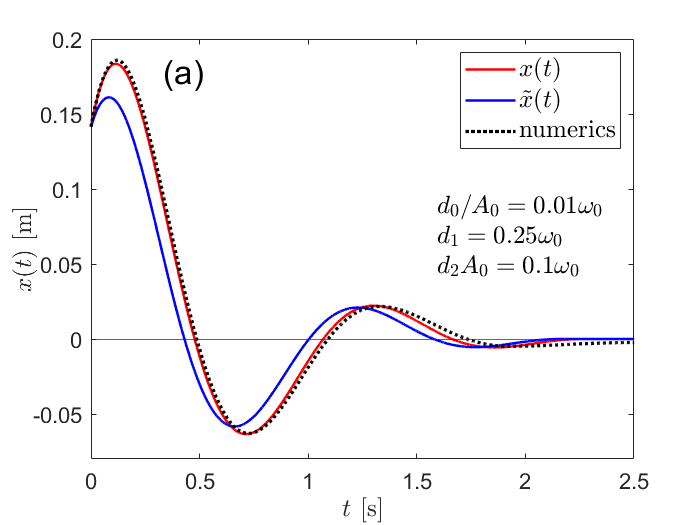}
\includegraphics[width=0.48\textwidth]{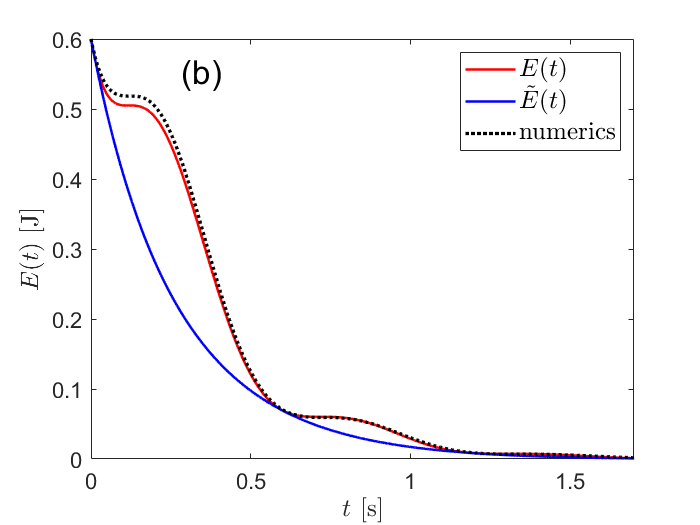}
\end{center}
\caption{(a) Solid red curve shows the solution $x(t)$ with initial conditions $(x_0=0.14m, v_0=0.77m/s)$ for $d_0/(\omega_0A_0)=0.01$, $d_1/\omega_0=0.25$ and $d_2A_0/\omega_0=0.1$. The solid blue curve and the black dotted curve show the corresponding $\tilde{x}(t)$ solution and the numerical solution. (b) Energies corresponding to solutions shown in figure (a). See text for details.} 
\label{slikaCLQ5}
\end{figure}

As an example, we take $x_0=A_0\sqrt{2}/2=0.14m$ and $v_0=\omega_0x_0=0.77m/s$. For this choice of initial conditions, the initial potential energy equals the initial kinetic energy, and $\varphi_0=-\pi/4$. Here we use (again) the \emph{roots} MATLAB function to solve the quartic equation \eqref{amplitudaLQC}. Thus, for the initial conditions $(x_0=0.14m,v_0=0.77m/s)$ with $d_0/(\omega_0A_0)=0.01$, $d_1/\omega_0=0.25$ and $d_2A_0/\omega_0=0.1$, we get $A=0.2464m$ and $\varphi=-0.9594$. The corresponding solutions $x(t)$ and $\tilde{x}(t)$, i.e. \eqref{xLQC} and \eqref{xLQC0}, are shown in Fig.\,\ref{slikaCLQ5}(a), and the corresponding energies are shown in Fig.\,\ref{slikaCLQ5}(b). We can see that $x(t)$ overlaps with the numerical solution significantly better than $\tilde{x}(t)$ up to the third turning point, after which $x(t)$ begins to deviate significantly from the numerical solution due to the dominant influence of Coulomb damping towards the end of the dynamics.

\subsection{A brief comment on the combination of quadratic and Coulomb damping and the combination of linear and Coulomb damping}
\label{brieflook}

For a harmonic oscillator damped by a combination of Coulomb and quadratic damping (i.e. the case with $d_0>0$, $d_2>0$, and $d_1=0$) one can easily obtain the corresponding approximate solutions simply by inserting $d_1=0$ and $\omega=\omega_0$ into expressions presented in this section. We analyze this combination of damping forces in detail in the first version of this work \cite{lelas2025newapproachapproximateanalytical}; however, here we omit a detailed analysis of this case to avoid overwhelming the reader, especially since all the necessary elements for such an analysis are presented in this section, allowing interested readers to conduct it themselves or they can consult \cite{lelas2025newapproachapproximateanalytical}.

For a harmonic oscillator damped by a combination of Coulomb damping and linear damping (i.e. the case with $d_0>0$, $d_1>0$, and $d_2=0$) the expressions presented in this section remain valid. Although the factor $1/d_2$ occurs in \eqref{rjes1}--\eqref{rjes3}, it does not occure in the unified expression \eqref{rjes11}: with $d_2=0$ we have $C=-d_1^2$, and \eqref{rjes11} reduces exactly to \eqref{fLC} below, while \eqref{tau1} reduces to \eqref{tetaLC}. As we noted in the introduction, this specific combination of damping allows exact analytical analysis, but exact solutions are piecewise defined, and simpler closed form approximate analytical solutions are still a subject of interest \cite{JSVCoulombViscous}. Therefore, we will briefly outline the solutions provided by our approach for this combination of damping forces in explicit form. In this case, \eqref{rjes11} (or, equivalently, solving the integral \eqref{Integral} directly) gives
\begin{equation}
f(t)=\frac{1}{d_1}\left[\left(\frac{d_0}{A}+d_1\right)e^{-d_1t}-\frac{d_0}{A}\right]\,.
\label{fLC}
\end{equation}
Thus, the corresponding approximate solution is
\begin{equation}
x(t)=\frac{A}{d_1}\left[\left(\frac{d_0}{A}+d_1\right)e^{-d_1t}-\frac{d_0}{A}\right]\cos(\omega t+\varphi)\,\Theta(\tau-t)\,,
\label{xLC}
\end{equation}
where $A$ and $\varphi$ are given by \eqref{amplitudaLQC} and \eqref{fazaLQC} evaluated with $d_2$ set to zero, while $\Theta(\tau-t)$ is given by \eqref{theta} with
\begin{equation}
\tau=\frac{1}{d_1}\ln\left(1+\frac{d_1A}{d_0}\right)\,.
\label{tetaLC}
\end{equation}
Since the primary goal of this work is to model cases that cannot be solved using exact analytical methods (due to quadratic damping), and since we are especially interested in the most realistic combination in which (to a greater or lesser extent) all three types of damping are present, we will not engage in a detailed analysis of the solution \eqref{xLC}.

\subsection{Comments on the results for a harmonic oscillator damped by all three types of damping}
\label{Clinearquadratic4}

The results in this section show that the new approximate analytical solutions \eqref{xLQC}, with the corresponding $A$ and $\varphi$, provide a much better description of the dynamics than the solutions \eqref{xLQC0} derived in \cite{LelasPezer2}. For the range of damping constants $\mu$, $b$, and $D$ such that $0<d_0/(\omega_0A_0)\lesssim0.03$, $0<d_1/\omega_0\lesssim0.25$, and $0<d_2A_0/\omega_0\lesssim0.25$ hold, solutions \eqref{xLQC} provide an excellent description of the dynamics for all types of initial conditions, unless we are concerned with high precision of the description at the very end of the dynamics, e.g. for the last or the last few half-periods before the halt.

As previously stated, these deviations towards the end of the dynamics arise because our approximate solutions, by their very form, cannot capture the increasing time shifts of the zero crossings or the halt at $x_{halt}\neq0$, i.e. the effects which arise due to Coulomb damping. One way to obtain an approximate solution that better describes the final stage of the dynamics is to neglect the contribution of quadratic damping during the last few half-periods before the halt (since its influence diminishes significantly as velocity decreases, and the velocity is low near the end of the motion anyway). In this way, we are left only with linear and Coulomb damping, a combination that allows for an exact analytical solution. We tested this approach for combined quadratic and Coulomb damping in the first version of this paper. It turned out that the final stage of the dynamics can be accurately described if we completely neglect quadratic damping, starting from the last turning point up to which the full approximate solution (which included both quadratic and Coulomb damping) performed well, take the displacement of that turning point as the initial condition, and solve exactly the equation of motion with Coulomb damping to describe the remaining dynamics \cite{lelas2025newapproachapproximateanalytical}. The resulting approximate solution has zero crossings and the halt position that coincide well with the numerically obtained values, but the price is a complicated procedure, and the solution is piecewise defined \cite{lelas2025newapproachapproximateanalytical}. Furthermore, in the next section, we show that solutions of the form \eqref{xLQC} can be used to determine damping constants from given displacement vs. time data, and a precise description of the dynamics right up to the very end is not required in this context. Therefore, here we omit the aforementioned approach to improving the modeling of the final stage of the dynamics.

\section{Potential application of our results in the analysis of the experimental data}
\label{application}

In this section, we present two simple examples demonstrating that our approximate solutions can be used to determine the damping constants from experimental data. For simplicity, we consider simulated data and employ a straightforward analysis using only basic MATLAB functions. However, analyzing experimental data with our solutions could be a topic of separate in-depth study. The choice of the most appropriate approach and fitting method depend on the characteristics and quality of the available data, as well as the required accuracy and the available computational tools. Thus, the examples presented here are intended primarily to illustrate the underlying idea rather than to provide a comprehensive treatment of the subject. 

\subsection{Analysis of a system with dominant linear and quadratic damping and barely detectable Coulomb damping}
\label{eksperiment1}

\begin{figure}[h!t!]
\begin{center}
\includegraphics[width=0.48\textwidth]{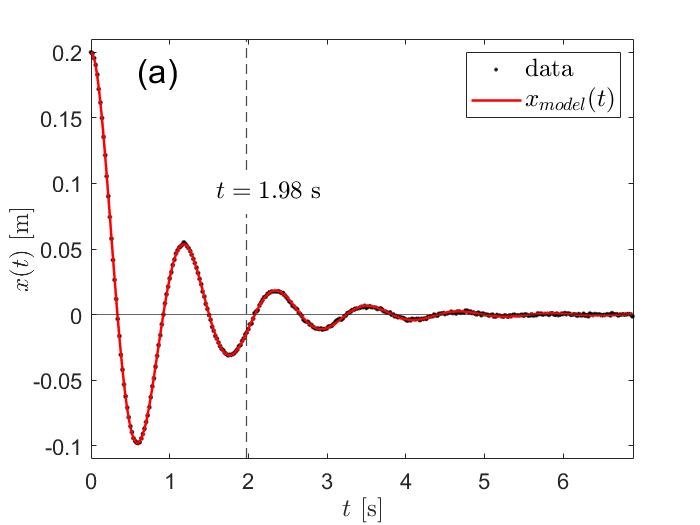}
\includegraphics[width=0.48\textwidth]{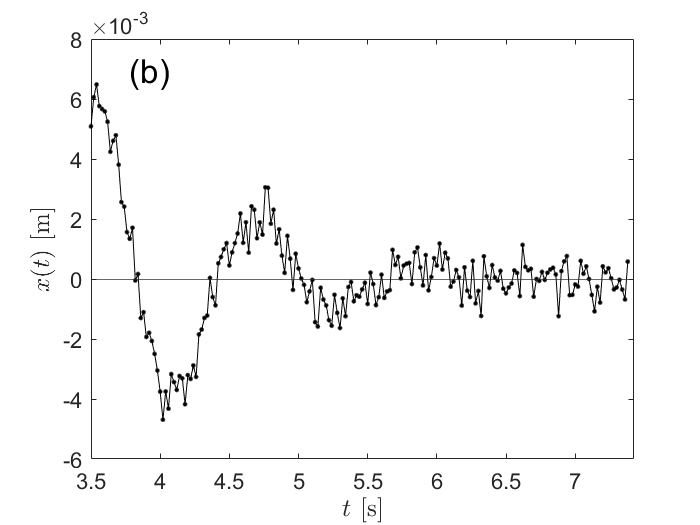}
\end{center}
\caption{(a) Simulated experimental data are shown as black dots. The solid red curve represents the optimal fit of solution \eqref{xLINKV} to the experimental data. The fitting was performed using only the data from time interval $0\leq t\leq1.98\,s$. (b) The simulated experimental data corresponding to the time interval $3.5\,s\leq t\leq 7.38\,s$. The data points are shown as black dots connected by lines to guide the eye. As can be seen, oscillations are no longer visible for $t\gtrsim6.3\,s$. We therefore take $t_{halt}=6.3\,s$ as an estimate of the halt instant. See text for details.} 
\label{slikaFIT1}
\end{figure}

In Fig.\,\ref{slikaFIT1}, we show simulated data (black dots) representing the oscillatory motion of our block-spring system with very low Coulomb damping, but moderately strong linear and quadratic damping. The data correspond to the numerical solution obtained with $(x_0=0.2m, v_0=0)$, $d_0/(\omega_0A_0)=5\cdot10^{-4}$ (i.e. $\mu=4.8\cdot10^{-4}$), $d_1/\omega_0=0.15$ (i.e. $b=1.643\,kg/s$) and $d_2x_0/\omega_0=0.1$ (i.e. $D=1.178\,kg/m$), to which Gaussian white noise has been added to simulate measurement errors. The noise was generated using \emph{randn} MATLAB function, with a standard deviation $\sigma=0.0025\,x_0$. The time interval between consecutive data points is $\Delta t=0.02\,s$. These values are within the capabilities of contemporary data acquisition systems (see, e.g., \cite{JSVLuan2023}).

When analyzing experimental data, we often have prior knowledge of certain fundamental properties of the system under investigation. For example, the data shown in Fig.\,\ref{slikaFIT1} could represent the motion of a glider that oscillates on an air track. In such a system, kinetic friction (Coulomb damping) exists but it is barely detectable \cite{Glider2026}, while the glider's motion may be significantly affected by linear \cite{MagnetDampingAirTrack} and quadratic damping \cite{AirTrackDrag}. The key point is that the underlying physics of a system can almost always reveal which factors dominate its dynamics and which are negligible or nearly negligible. Accordingly, the analysis of the data presented in Fig.\,\ref{slikaFIT1} can begin with the understanding that we are dealing with harmonic oscillations subject to moderately strong linear and quadratic damping, whereas Coulomb damping is considerably weaker.

Therefore, the following strategy can be adopted: rather than using the approximate solution \eqref{xLQC} (obtained for the combination of three types of damping), we use the approximate solution \eqref{xLINKV} (obtained for the combination of linear and quadratic damping) to model the simulated experimental data. This approach allows us to work with a mathematically simpler function, while acknowledging that neglecting weak Coulomb damping may introduce errors when the optimization includes data from later stages of the motion, particularly the final stage. 

For this reason, the fitting procedure was restricted to the first $100$ data points from the simulated experiment, corresponding to the time interval $0\leq t\leq1.98\,s$ (i.e. the time interval indicated by the vertical dashed line in Fig.\,\ref{slikaFIT1}(a)). We fit the data to function \eqref{xLINKV} using $A$, $\omega$, $d_1$, and $d_2$ as free parameters. Rather than treating the phase $\varphi$ of \eqref{xLINKV} as an independent free parameter, we expressed it in terms of the amplitude parameter as $\varphi=-\arccos(x_0/A)$, where $x_0$ is the known initial displacement, ensuring that the model reproduces the prescribed initial displacement.

The optimal model parameters were determined using a fitting procedure based on \emph{fminsearch} MATLAB function. Broad bounds were assigned to the free parameters, with the following lower and upper limits:
\begin{equation}
x_0\leq A\leq2x_0\,,\,0.5\omega_0\leq\omega\leq1.5\omega_0\,,\,0\leq d_1\leq0.5\omega_0\,,\,0\leq d_2\leq0.5\omega_0/x_0\,. 
\label{granice}
\end{equation}
Here, we consider a block-spring system with known $\omega_0$, but note that establishing such bounds on the free parameters does not require prior knowledge of $\omega_0$. A rough estimate of the frequency $\omega$ can easily be obtained directly from the data, for example, by examining the time intervals between successive zero crossings or turning points. This estimated frequency can then be used to establish bounds on the parameters $\omega$, $d_1$, and $d_2$ as in the relation \eqref{granice}. We choose the initial parameter values as the midpoints of these intervals. As the optimal values of the free parameters, we obtain:
\begin{equation}
A_{opt}=0.206\,m\,,\,\omega_{opt}=5.397\,s^{-1}\,,\,d_{1opt}=0.833\,s^{-1}\,,\,d_{2opt}=2.716\,m^{-1}s^{-1}\,. 
\label{optimalni1}
\end{equation}
Now, using $d_{1opt}$ and $d_{2opt}$ in the corresponding two expressions given in \eqref{coeffC}, we get
\begin{equation}
b_{opt}=1.667\,kg/s\,\,\,\text{and}\,\,\,D_{opt}=1.169\,kg/m\, 
\label{optimalni2}
\end{equation}
as the optimal linear and quadratic damping constants determined from the simulated experimental data. If, for example, we had considered a system for which the exact value of $\omega_0$ was unknown, we could determine it as $\sqrt{\omega_{opt}^2+d_{1opt}^2}$ and use that value to find the constant $D_{opt}$.

We can use the values $A_{opt}$, $d_{1opt}$, and $d_{2opt}$, given in \eqref{optimalni1}, to estimate $d_0$, and then $\mu$. We approximate the relation \eqref{tau1} with
\begin{equation}
\frac{2}{\sqrt{4d_{2opt}\,d_0-d_{1opt}}}\,\arctan\left(\frac{A_{opt}\sqrt{4d_{2opt}\,d_0-d_{1opt}}}{2d_0+d_{1opt}\,A_{opt}}\right)-t_{halt}=0\,,
\label{tauEXP}
\end{equation}
where $t_{halt}=6.3\,s$ is the halt instant estimated from the data (see Fig.\,\ref{slikaFIT1}(b)). We solve \eqref{tauEXP} with $d_0$ as a variable, using \emph{fzero} MATLAB function, and get $d_0=5.52\cdot10^{-4}\,m/s$. Using this value and the connection of $d_0$ and $\mu$ given in \eqref{coeffC}, we get $\mu_{opt}=4.84\cdot10^{-4}$.

As noted at the beginning of this subsection, the simulated experimental data were created with $\mu_{d}=4.8\cdot10^{-4}$, $b_{d}=1.643\,kg/s$ and $D_{d}=1.178\,kg/m$, where we use subscript $d$ simply to denote these values. Thus, the relative errors of our "experimentally" determined damping constants are
\begin{equation}
R_b=1.46\%\,\,,\,\,R_D=0.76\%\,\,,\,\,R_{\mu}=0.84\%\,,
\label{error}
\end{equation}
where $R_b=|b_{opt}-b_{d}|/b_{d}\cdot100\%$, $R_D=|D_{opt}-D_{d}|/D_{d}\cdot100\%$, $R_{\mu}=|\mu_{opt}-\mu_{d}|/\mu_{d}\cdot100\%$. Therefore, a relatively simple analysis was sufficient to obtain satisfactory results. Furthermore, we have checked that the results are not significantly affected by different realizations of Gaussian noise. In particular, even a twofold increase in the standard deviation of the noise does not lead to any substantial changes in the estimated parameters, with the corresponding relative errors remaining of the same order of magnitude. We also verified that using a somewhat longer or shorter time interval, corresponding to about $50$ more or fewer data points, does not significantly affect the results. We also verified that estimates of the halt instant within the range $6.2s\leq t_{halt}\leq6.4s$ correspond to values of $\mu_{opt}$ with $R_{\mu}<10\%$. Overall, these results demonstrate the robustness of the presented analysis. 

We note here that since $\omega$ and $A$ were treated as free parameters in the fitting procedure, an additional check of the agreement between the simulated experimental data and our theoretical predictions can be performed by comparing the values of $\omega_{opt}$, $A_{opt}$ and $\varphi_{opt}=-\arccos(x_0/A_{opt})$ with the values obtained by substituting $d_{1opt}$ and $d_{2opt}$ into theoretical expressions for the frequency, initial amplitude, and initial phase of the model solution considered, i.e. of \eqref{xLINKV}. Thus, the expressions for the frequency is $\omega(d_1)=\sqrt{\omega_0^2-d_1^2}$, while the initial amplitude $A(d_1,d_2)$ and the initial phase $\varphi(d_1,d_2)$ are given in \eqref{AF1LINKV}. We get $\omega(d_{1opt})=5.414\,s^{-1}$, $A(d_{1opt},d_{2opt})=0.207\,m$ and $\varphi(d_{1opt},d_{2opt})=-0.252$. The relative errors of these values with respect to $\omega_{opt}$, $A_{opt}$ and $\varphi_{opt}$ are $0.32\%$, $0.49\%$ and $4.13\%$, respectively. Therefore, the overall agreement between the theory and the data is excellent.

\subsection{Analysis of a system with unknown linear and quadratic damping and known Coulomb damping}
\label{eksperiment2}

\begin{figure}[h!t!]
\begin{center}
\includegraphics[width=0.48\textwidth]{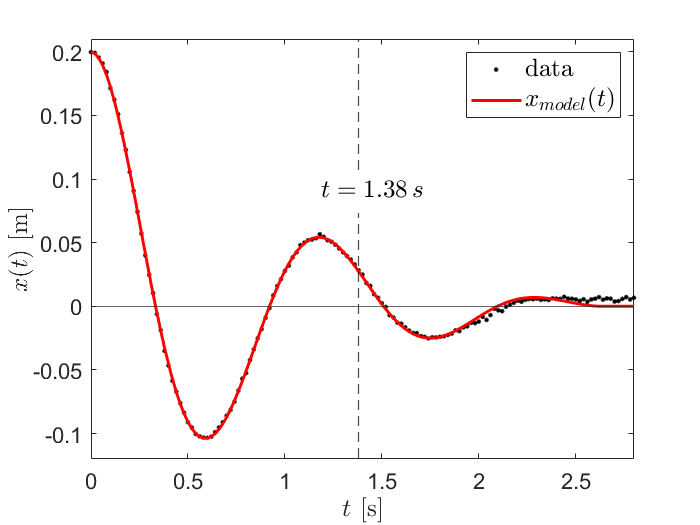}
\end{center}
\caption{Simulated experimental data are shown as black dots. The solid red curve represents the optimal fit of solution \eqref{xLQC} to the experimental data. The fitting was performed using only the data from time interval $0\leq t\leq1.38\,s$. See text for details.} 
\label{slikaFIT2}
\end{figure}

In Fig.\,\ref{slikaFIT2}, we show simulated data (black dots) representing the oscillatory motion of a block–spring system in which Coulomb damping plays a significant role throughout the motion, together with linear and quadratic damping. The data correspond to the numerical solution obtained with $(x_0=0.2m, v_0=0)$, $d_0/(\omega_0A_0)=0.02$ (i.e. $\mu=0.019$), $d_1/\omega_0=0.12$ (i.e. $b=1.315\,kg/s$) and $d_2x_0/\omega_0=0.08$ (i.e. $D=0.943\,kg/m$), to which Gaussian white noise with $\sigma=0.005\,x_0$ has been added. The time interval between consecutive data points is again $\Delta t=0.02\,s$.

Given the significant contribution of Coulomb damping, its dynamic friction coefficient $\mu=0.019$ is assumed to be known, as it can be determined independently through non-oscillatory experiments, such as constant-velocity measurements in which linear and quadratic damping effects can be neglected for sufficiently small velocities (see, e.g., \cite{DeAmbrosis}). Therefore, we use the approximate solution \eqref{xLQC} (obtained for the combination of three types of damping), again with $A$, $\omega$, $d_1$ and $d_2$ as free parameters (and $\varphi=-\arccos(x_0/A)$), to model the simulated experimental data. The fitting procedure was restricted to the first $70$ data points from the simulated experiment, corresponding to the time interval $0\leq t\leq1.38\,s$ (i.e. the time interval indicated by the vertical dashed line in Fig.\,\ref{slikaFIT2}). We selected this interval to avoid data corresponding to the last two zero crossings and the final stage of the motion, where the model is known to become less accurate.  

Here, we use the same parameter bounds, MATLAB functions, and optimization procedure as described in the previous subsection, the only difference is that we use the approximate solution \eqref{xLQC} as the model. We get
\begin{equation}
A_{opt}=0.205\,m\,,\,\omega_{opt}=5.346\,s^{-1}\,,\,d_{1opt}=0.652\,s^{-1}\,,\,d_{2opt}=2.090\,m^{-1}s^{-1}\,, 
\label{optimalni22}
\end{equation}
and 
\begin{equation}
b_{opt}=1.303\,kg/s\,\,,\,\,D_{opt}=0.899\,kg/m\, 
\label{optimalni33}
\end{equation}
as the corresponding linear and quadratic damping constants. Thus, the relative errors are
\begin{equation}
R_b=0.86\%\,\,,\,\,R_D=4.61\%\,.
\label{error2}
\end{equation}
Again, a relatively simple analysis was sufficient to obtain solid results.

\section{Conclusion and outlook}
\label{Conclusion}

By comparing the new approximate analytical solutions with numerical solutions, we demonstrated their excellent agreement in describing the free vibrations of a harmonic oscillator subject to purely quadratic damping, combined linear and quadratic damping, and combined linear, quadratic, and Coulomb damping. The proposed approach is applicable to arbitrary initial conditions and a broad range of damping parameters, making it a versatile tool for modeling and analyzing free damped vibrations of single-degree-of-freedom systems, with potential applications in experimental physics and engineering. Throughout this paper, we have used the block-spring system as a paradigmatic example of a damped oscillatory system. However, the presented analysis can easily be adapted to any system described by the damped harmonic oscillator model. 


From a practical point of view, we have demonstrated, through two simple examples, that our results can be used to estimate the damping constants from experimentally measured free-vibration responses. In particular, for a single degree of freedom system, we can excite free vibrations with the initial conditions $(x_0>0,v_0=0)$, measure the displacement as a function of time, and then fit the experimental data to a function having the same form as one of our approximate analytical solutions (the one that is most appropriate for the experiment under consideration), with the damping constants treated as fitting parameters. We can avoid the limitations of our solutions in describing the later stages of motion of systems with Coulomb damping by excluding, for example, the last few half-periods from the fitting procedure.

We emphasize that the approach introduced in this paper can be readily extended to other combinations of linear and nonlinear damping not considered here. For example, the combination of linear and cubic damping is of interest in engineering (see, e.g., \cite{JSVMeskell2006, JSVLang2009}), and our approach can be easily adapted to this particular case. Furthermore, in the present work, we briefly discussed how to use our solutions to analyze experimental data. In future work, we intend to investigate the application of the proposed approach to other combinations of linear and nonlinear damping forces, and to explore in detail how the newly derived solutions can be used most effectively to extract information about damping from experimental data.

\section*{Acknowledgments}

K.L. and R.P. acknowledge support from the project “Implementation of cutting-edge research and its application as part of the Scientific Center of Excellence for Quantum and Complex Systems, and Representations of Lie Algebras”, Grant No. PK.1.1.10.0004, co-financed by the European Union through the European Regional Development Fund - Competitiveness and Cohesion Programme 2021-2027. In addition, R.P. acknowledge support from the European Regional Development Fund, Operational Programme Competitiveness and Cohesion 2014–2020 (Project name Centar za ljevarstvo-SIMET, Code KK.01.1.1.02.0020).





\bibliographystyle{elsarticle-num}  
\bibliography{ZadnjaVerzija3}

\end{document}